\documentclass[a4paper,11pt]{article}
\usepackage{jcappub} % for details on the use of the package, please see the JINST-author-manual
\usepackage{aas_macros} % for details on the use of the package, please see the JINST-author-manual
\usepackage{lineno}

\usepackage{graphicx}	% Including figure files
\usepackage{amsmath}	% Advanced maths commands

\def\bi#1{\hbox{\boldmath{$#1$}}}

\usepackage[normalem]{ulem}

\usepackage[T1]{fontenc} % if needed

\usepackage{url}

\title{\boldmath DeepWiener: Neural Networks for CMB polarization maps and power spectrum computation}

\author[a,b,1]{Bel\'en Costanza,\note{Corresponding author.}}
\author[c]{Claudia G. Sc\'occola,}
\author[d]{Mat{\'\i}as Zaldarriaga}

\affiliation[a]{ Facultad de Ciencias Astron\'omicas y Geof\'isicas, Universidad Nacional de La Plata, Observatorio Astron\'omico, Paseo del Bosque, \\
B1900FWA La Plata, Argentina }
\affiliation[b]{Consejo Nacional de Investigaciones Cient\'ificas y T\'ecnicas (CONICET), Rivadavia 1917, Buenos Aires, Argentina }
\affiliation[c]{Departamento de F\'isica, FCFM, Universidad de Chile, Blanco Encalada 2008, Santiago, Chile}
\affiliation[d]{School of Natural Sciences, Institute for Advanced Study, 1 Einstein Drive, Princeton, NJ 08540, USA}

\emailAdd{belen@fcaglp.unlp.edu.ar}
\emailAdd{claudia.scoccola@uchile.cl}
\emailAdd{matiasz@ias.edu}

\abstract{To study the early Universe, it is essential to estimate cosmological parameters with high accuracy, which depends on the optimal reconstruction of Cosmic Microwave Background (CMB) maps and the measurement of their power spectrum. In this paper, we generalize the neural network developed for applying the Wiener Filter, initially presented for temperature maps in previous work, to polarization maps. Our neural network has a UNet architecture, including an extra channel for the noise variance map, to account for inhomogeneous noise, and a channel for the mask. In addition, we propose an iterative approach for reconstructing the E and B-mode fields, while addressing the E-to-B leakage present in the maps due to incomplete sky coverage. The accuracy achieved is satisfactory compared to the Wiener Filter solution computed with the standard Conjugate Gradient method, and it is highly efficient, enabling the computation of the power spectrum of an unknown signal using the optimal quadratic estimator. We further evaluate the quality of the reconstructed maps at the power spectrum level along with their corresponding errors, finding that these errors are smaller than those obtained using the well-known pseudo-$C_\ell$ approach. Our results show that increasing complexity in the applied mask presents a more significant challenge for B-mode reconstruction.}

\begin{document}

\maketitle

\flushbottom

%% En el archivo paper_relacionados, hay algunos comentarios sobre papers que nos pueden servir en el futuro. Lo comento por ahora.
%%\input{papers_relacionados}
 
\section{Introduction}
\label{sec:introduction}

%blabla del CMB polarization, de las redes neuronales, del trabajo anterior, y de las cotas sobre B modes.

%En este trabajo vamos a lidiar con el E-to-B leakage. Y vamos a explicar la metodología que implementamos para minimizar el leakage. Por alguna razón no era necesario hacerla en TensorFlow1, pero así es la vida.
%(En un follow-up paper vamos a aplicar esto a los mapas de QUBIC, analizando el ruido de los mismos. Ese seguramente sea un paper de colaboración).

The Cosmic Microwave Background (CMB) is one of the richest sources of information for understanding the origin and evolution of our Universe. This relic radiation was released soon after the recombination epoch, when the Universe was 380,000 years old. Precise measurements of the CMB temperature and polarization anisotropies allow for accurate estimation of the cosmological parameters, helping to constrain current cosmological models. 

Primordial metric perturbations can be split into scalar and tensor types, simplifying the equations for the evolution of the perturbations in matter and in radiation temperature and polarization. While both scalar and tensor perturbations source temperature anisotropies, CMB polarization shows a distinct behavior: decomposing polarization into gradient (E-modes) and curl (B-modes) components reveals that E-modes receive contributions from both scalar and tensor perturbations, whereas B-modes arise exclusively from primordial gravitational waves (tensor perturbations)~\citep{1997PhRvD..55.1830Z}.
The CMB polarization pattern has been measured by satellite-based experiments like, WMAP~\citep{2013ApJS..208...19H,2013ApJS..208...20B} and Planck~\citep{2020A&A...641A...1P,2020A&A...641A...5P}, as well as ground-based experiments, like ACT~\citep{2020JCAP...12..047A, 2020JCAP...12..045C} and SPTpol~\citep{2018ApJ...852...97H}, specifically the E-mode polarization with high accuracy.  

%The temperature of the CMB includes contributions from the scalar modes (density perturbations) and tensor modes, while the polarization can be decomposed into gradient and curl components, referred to as E-modes and B-modes. The E-mode polarization is generated by scalar density perturbations, whereas B-mode polarization is produced by the primordial gravitational waves (tensor perturbations)~\citep{1997PhRvD..55.1830Z}. The CMB polarization pattern has been measured by satellite-based experiments like, WMAP~\citep{2013ApJS..208...19H,2013ApJS..208...20B} and Planck~\citep{2020A&A...641A...1P,2020A&A...641A...5P}, as well as ground-based experiments, like ACT~\citep{2020JCAP...12..047A, 2020JCAP...12..045C} and SPT~\citep{2018ApJ...852...97H}, specifically the E-mode polarization with high accuracy.  

Measurements of B-mode polarization remain challenging yet are crucial, as they provide direct information about the stochastic background of gravitational waves left over from the inflationary epoch. This constitutes a major goal of cosmology, enabling constraints on cosmological models and offering insight into the inflationary energy scale.
Future and on-going experiments, like Lite-Bird~\citep{2020SPIE11443E..2FH}, PICO~\citep{2019arXiv190210541H}, SPT-3G~\citep{2014SPIE.9153E..1PB}, Simons Observatory~\citep{2019JCAP...02..056A}, CMB-S4~\citep{2019arXiv190704473A}, and QUBIC~\citep{Hamilton2022, 2022JCAP...04..035M, 2018arXiv181200785M}, with a large number of highly sensitive detectors, attempt to measure the B-mode polarization.

The emergence of new experiments with ambitious technological requirements, together with the increasing volume of CMB data they will provide, demands the development of sophisticated tools to maximize the extraction of cosmological information. There are three essential data compression stages in a CMB pipeline: the conversion of raw time-ordered data (TOD) into a sky map (map making), the estimation of the power spectra, and the inference of cosmological parameters~\citep{1997PhRvD..55.5895T}. At each stage, the dataset is compressed to a more manageable size. To ensure that no significant cosmological information is lost, every step must be carefully optimized for both accuracy and computational efficiency.

In this context, we address the challenge of optimal reconstruction of polarization maps and power spectrum estimation using noisy and sparse data. The well-known Wiener Filter (WF)~\citep{Wiener1949} is the optimal filter for noise reduction in these maps and for reconstructing the underlying signal~\citep{1995ApJ...449..446Z}, as it minimizes the residual variances in the reconstruction of the true field from noisy observations. For the CMB, which is a Gaussian distributed field, the WF solution maximizes the posterior probability of the signal given the data~\citep{Seljak:1997wx}. 

However, the computational demands for deriving the exact WF solution have increased significantly due to the vast amount of data involved. Additionally, calculating the WF matrix requires inverting the covariance matrices of both signal and noise, where the signal covariance is diagonal in Fourier space, while the noise covariance is diagonal in configuration space. Then, the combined covariance is neither diagonal nor sparse in any basis. One approach to solving the WF is to employ a conjugate gradient method for matrix inversion with an effective preconditioner~\citep{10.5555/1403886}, which is a complicated task as it must approximate the true inverse while maintaining sparsity. Examples of preconditioners include the inverse of the diagonal part of the matrix~\citep{2004PhRvD..70h3511W} or the multigrid preconditioner~\citep{2007PhRvD..76d3510S}, among others. Other methods without preconditioners include the incorporation of a messenger field between bases~\citep{2013A&A...549A.111E, 2017MNRAS.468.1782K}. 

For this reason, in this paper, we attempt to simulate the WF considering Deep Learning techniques, since neural networks have proven effective in various cosmological analysis~\citep{2019MNRAS.484.5771A, 2020ApJ...903..104P}. After that, we proceed to estimate the power spectrum, as we have done on temperature maps~\citep{2024JCAP...04..041C}, but generalized to polarization, implementing an optimal quadratic estimator approach suggested by~\citep{2017JCAP...12..009S, 2019JCAP...10..035H}, obtained from the maximization of the likelihood. The use of neural networks to simulate the WF allows to compute a simulation-based optimal quadratic estimator, as the trained models are more efficient for performing the WF compared to the Conjugate Gradient method, which can become prohibitively expensive in certain cases.

Due to foreground contamination from the galactic plane, the local Universe, and extragalactic sources in CMB observations, experiments typically measure a partial CMB sky by applying a binary mask to the full-sky map. One complication that arises from incomplete sky coverage is the transmission of power from the E-mode to the B-mode, a problem commonly referred to as E-to-B leakage. Additionally, since the B-component is expected to be at least an order of magnitude smaller than the E-component, a clean separation of the modes is necessary.
Several approaches have tried to mitigate the E-to-B leakage in partial sky analysis. Some of these focus on separating the contributions of the E and B \textit{pure} components from the \textit{ambiguous} modes, which cause the E-to-B leakage, by finding the orthonormal bases for these three components~\citep{2003PhRvD..67b3501B}, or by using the pure pseudo-$C_{\ell}$ formalism~\citep{2007PhRvD..76d3001S}. Other approaches attempt to mitigate the E-to-B leakage at the power spectrum level using deep learning techniques~\citep{2023JApA...44...84P}. 

In this work, we address the challenge of applying the WF to polarization maps while simultaneously dealing with E-to-B leakage. We start by obtaining the WF for the E-mode, as the Q and U maps predominantly consist of E-modes due to their stronger signal. We then create a new dataset by removing the contribution of the E-mode WF, allowing the network to focus on the B-modes.  This approach is motivated by~\citep{2017PhRvD..96d3523B} where the modes decomposition is considered as an application of the WF. If one can identify and remove the E-mode contribution from the data, it will not interfere with the WF of the B-mode. 

%[Claudia: modifico esto para que quede claro que el inhomogeneous noise ya era una mejora nuestra respecto de Wienernet] We use the neural network already implemented on temperature maps for inhomogeneous noise, with an additional channel added, based on a neural network built for homogeneous noise called WienerNet~\citep{2019arXiv190505846M}. Since it has been proved in our previous work that the WF predictions, after the training of the models, is quite efficient~\citep{2024JCAP...04..041C}, we proceed to estimate the power spectrum amplitudes of the polarization field and their covariance matrix. We have written our codes in \textsc{tensorflow} 2 and \textsc{keras}, that are based on \textsc{python} 3. 

We have adapted the WienerNet neural network~\citep{2019arXiv190505846M}, originally developed for homogeneous noise, to handle inhomogeneous noise in temperature maps by incorporating an additional channel with an inhomogeneous variance map, as introduced in our previous work~\citep{2024JCAP...04..041C}. As demonstrated in that work, the WF predictions after model training are highly efficient. In this work, we proceed to estimate the wiener filtered map and the power spectrum of the polarization field and its covariance matrix. Our code is implemented in \textsc{Tensorflow} 2 and \textsc{Keras}, using \textsc{Python} 3.

In this paper, we focus on polarization maps and present a method for reconstructing the E- and B-mode fields, accounting for realistic inhomogeneous noise and varying coverage complexities. Our aim is to provide a practical framework for current and future experiments, as the Q $\&$ U Bolometric Interferometer for Cosmology (QUBIC) experiment, which has a multi-peak synthesized beam~\citep{2018arXiv181200785M} that lead to spectro-imaging with spatial and subfrequency correlations.

This paper is organized as follows: In section~\ref{sec:methodology} we described the basics of Wiener filter and power spectrum estimation, with the loss functions implemented in this work. In section~\ref{sec:deepwiener} we present the neural network architecture developed for our purposes called DeepWiener. In section~\ref{sec:datasets} we explain how we created the necessary datasets with the different masks applied and noise properties. Then, in section~\ref{sec:results} we presents the neural network results of the E and B-modes compared with the WF using a conjugate gradient method, together with the power spectrum estimation of both modes. In addition, in the same section we compare our framework with the pseudo-$C_{\ell}$ approach. The discussion and conclusions are presented in section~\ref{sec:conclusion}. 
%\textcolor{magenta}{falta explicar más sobre power spectrum o pseudo-cl? más adelante lo hago}

\section{Methodology}
\label{sec:methodology}

%Resumen de WF y del power spectrum estimation, teniendo en cuenta que esto es un follow-up paper. Puede estar mas resumido

%\input{methodology}

In this section, we introduce the Wiener filter and the methodology adopted to implement it using neural networks. Additionally, we provide an overview of the algorithm used to estimate the power spectra of CMB polarization maps (E-mode power spectrum and B-mode power spectrum) after applying the Wiener filter.

\subsection{Wiener Filter and loss function}

The Wiener Filter is the optimal filter to enhance signal quality by reducing the noise present on gaussian fields, as in the case of the CMB signal. Moreover, for CMB observations, the data is a linear transformation of the initial modes. Hence, let us suppose that the measurements \textbf{d} are a linear combination of the underlying field \textbf{s} that we want to estimate:
\begin{equation}
    \textbf{d} = \textbf{R} \textbf{s} + \epsilon, 
    \label{eq:eq1}
\end{equation}
where \textbf{R} is the response matrix of the measurement procedure and $\epsilon$ is the data uncertainty.

The application of a filter, mathematically expressed as a convolution, establishes a linear connection between the reconstructed underlying signal $\hat{\textbf{s}}_{WF}$ and the input data \textbf{d}, such that $\hat{\textbf{s}}_{WF} = M \textbf{d}$. The matrix M is determined by minimizing the variance of the residuals between the reconstructed signal and the true underlying signal.
In the case of Gaussian random fields, this result coincides with the Bayesian estimator that maximizes the conditional probability of the signal given the data: 
\begin{equation}
 P(\textbf{s} |\textbf{d})  \propto 
     \exp \left[-\frac{1}{2}(\textbf{s}^{\dag}\textbf{S}^{-1}\textbf{s}+ (\textbf{d}-\textbf{R}\textbf{s})^{\dag}\textbf{N}^{-1}(\textbf{d}-\textbf{R}\textbf{s}))\right],
   \label{eq:eq16}                     
\end{equation}
where $\textbf{S}$ and $\textbf{N}$ are the covariance matrices of the signal and noise, respectively. 
Maximizing this conditional probability is equivalent to minimizing the function: 
\begin{equation}
    \chi^{2}(\textbf{s}) = \textbf{s}^{\dagger} \textbf{S}^{-1} \textbf{s} + (\textbf{d}-\textbf{R}\textbf{s})^{\dagger} \textbf{N}^{-1} (\textbf{d}-\textbf{R}\textbf{s}).
\label{chi}
\end{equation}
%
%The corresponding result of minimizing~\ref{chi} is the Wiener Filter estimator: 
which results in the Wiener Filter estimator: 
\begin{equation}
\hat{\textbf{s}}_{WF} = \textbf{S}(\textbf{S}+\textbf{N})^{-1}\textbf{R}^{-1} \textbf{d} = 
 (\textbf{S}^{-1} + \textbf{R}^{\dagger} \textbf{N}^{-1} \textbf{R})^{-1} \textbf{R}^{\dagger} \textbf{N}^{-1} \textbf{d}. 
\label{eq:eqWF}
\end{equation}
%
%\textcolor{red}{[Claudia: hay algo que no entiendo acá: como pasas de la expresion del medio de (2.4) a la expresion de la izquierda de (2.4) ??]}\textcolor{magenta}{son la misma matriz, hay que trabajar un poquito la expresion derecha y llegas a la de la izquierda. s = WF d}
%
%To obtain the exact 
To compute this exact WF estimator, the expression~\eqref{eq:eqWF} is written as a linear system and inverted through the Preconditioner Conjugate Gradient (PCG) algorithm. 
In this work, we compare the performance of the neural network predictions with the WF results obtained using the PCG method. For detailed information about the PCG algorithm, refer to Appendix~\ref{apx1}.

For training a neural network to simulate the WF, it is natural to choose the same expression as in equation~\eqref{chi} for the loss function:
%For the training of a neural network that aims to simulate the WF is natural to choose as a loss function the same expression as~\eqref{chi}: 
%
\begin{equation}
J(d,y) = \frac{1}{2} (y-d)^{T}N^{-1}(y-d) + \frac{1}{2}y^{T}S^{-1}y,
\end{equation}
where $y$ is the output of the neural network that represents the reconstructed signal, $d$ is the noisy data ($Q_{obs}$ and $U_{obs}$), and $N$ and $S$ are the covariance matrices associated with noise and signal, respectively. The implementation for polarization maps is:
\begin{equation}
    J_{Q,U} = \sum_{i}^{N_{pix}} \frac{\left(Q_{NN} - Q_{obs}\right)^{2}}{\sigma_{i}^{2}} + \frac{\left(U_{NN} - U_{obs}\right)^{2}}{\sigma_{i}^{2}} + \sum_{l} \frac{E_{lNN}E_{lNN}^*}{C_{l}^E}  + \frac{B_{lNN}B_{lNN}^*}{C_{l}^B},
\label{pol}
\end{equation}
where $Q_{obs}$ and $U_{obs}$ are the inputs of the neural network, $Q_{NN}$ and $U_{NN}$ are the predictions, $\sigma^{2}_{i}$ is the pixel noise variance, and the Fourier terms are evaluated in the ($E$,$B$) basis. These expressions are the same as the ones presented in~\citep{2019arXiv190505846M}.

%We aim to obtain the $E_{WF}$ and $B_{WF}$ that are contained in $Q_{WF}$ and $U_{WF}$. 
We aim to extract the $E_{WF}$ and $B_{WF}$ components from $Q_{WF}$ and $U_{WF}$. 
In order to address the challenge of the E-to-B leakage, we performed the optimization of these components separately and iteratively, in a series of steps. 

First, we built the observed polarization maps $Q_{obs}^{(1)}$ and $U_{obs}^{(1)}$ with inhomogeneous noise and mask applied. Then, we trained the neural network and obtained $Q_{NN}^{(1)}$ and $U_{NN}^{(1)}$ as outputs, which gave us $E_{NN}^{(1)}$ and $B_{NN}^{(1)}$ after the appropriate transformation in Fourier space for the partial sky, which is explained in Appendix~\ref{apx2}. Since the B-mode is a weak signal compared to the E-mode, the neural network primarily captures the Wiener Filter of the E-mode but almost nothing of the B-mode. 

%\textcolor{red}{When the loss function $J_{Q,U}$, in equation~\eqref{pol}, is correctly minimized the outputs of the neural network should be the WF of $Q$ and $U$, which are then transformed to $E_{WF}$ and $B_{WF}$. In practice, the optimization of the neural network will provide an approximation to the WF. Specifically, in our case, in the first training we are able to reconstruct an approximated E-mode contribution but poorly the B-mode, being not able to accurately isolate the \textit{ambiguous} modes from the \textit{pure} ones.} 

When the loss function $J_{Q,U}$ in equation~\eqref{pol} is correctly minimized, the outputs of the neural network should correspond to the WF of $Q$ and $U$, which are then transformed into $E_{WF}$ and $B_{WF}$. In practice, however, the optimization of the neural network provides an approximation to the WF. Specifically, during the initial training, we can accurately reconstruct the E-mode contribution, but struggle with the B-mode, as the network fails to accurately isolate the \textit{ambiguous} modes from the \textit{pure} ones.

Several implementations could affect the optimization of a neural network, we do not discard the possibility of finding even better results by expanding the set of hyperparameters tested or considering different initializations, at the expense of more computing time for training. In addition, this architecture is written in an updated version of \textsc{Tensorflow} in order to be used in \textsc{Python 3} and with the new GPUs drivers. The WienerNet network (in which our network is based) was written in \textsc{Tensorflow 1} and the B-modes were obtained at the first training, indicating some differences in the internal implementations of the \textsc{Tensorflow} library that could affect the search of an optimal solution.

Therefore, we propose a method to extract as much of the E-mode contribution from the data as possible, removing it afterward to avoid contaminating the WF for the B-mode. Subsequently, we generate a new dataset with the E-mode contribution removed: 
\begin{align}
    Q^{(2)}_{obs} = & \left(Q_{obs} - Q_{ENN}^{(1)}\right) \\
    U^{(2)}_{obs} = & \left(U_{obs} - U_{ENN}^{(1)}\right),
\label{QU1}
\end{align}
where $Q_{ENN}^{(1)}$ and $U_{ENN}^{(1)}$ are the E-mode contributions of the first neural network outputs. This new dataset has a lower contribution of the E-mode in $Q^{(2)}_{obs}$ and $U^{(2)}_{obs}$, reducing the leakage present in the B-mode. 

We trained the neural network again with these residuals using the loss function $J_{Q,U}$, equation~\eqref{pol}, which now takes the form:  
\begin{equation}
    J_{Q,U} = \sum_{i}^{N_{pix}} \frac{\left(Q^{(2)}_{NN} - Q^{(2)}_{obs}\right)^{2}}{\sigma_{i}^{2}} + \frac{\left(U^{(2)}_{NN} - U^{(2)}_{obs}\right)^{2}}{\sigma_{i}^{2}} + \sum_{l} \frac{E_{lNN}E_{lNN}^*}{C_{l}^E}  + \frac{B_{lNN}B_{lNN}^*}{C_{l}^B},
\label{pol2}
\end{equation}
%
%where we added in the Fourier E-mode term the contribution $E_{NN}^{(1)}$ since the $C_{l}^{E}$ is the total power spectrum. 
where we added the contribution $E_{NN}^{(1)}$ to the Fourier E-mode term, as $C_{l}^{E}$ represents the total power spectrum.
%After that, we obtained as outputs $Q_{NN}^{(2)}$ and $U_{NN}^{(2)}$. Since $Q^{(2)}_{obs} = Q_{obs} - Q_{ENN}^{(1)}$, hence, $Q_{NN} = Q^{(2)}_{NN}+Q^{(1)}_{ENN}$ and $(Q^{(2)}_{NN} - Q^{(2)}_{obs}) = (Q_{NN} - Q_{obs})$.
Afterward, we obtained $Q_{NN}^{(2)}$ and $U_{NN}^{(2)}$ as outputs. Since $ Q^{(2)}_{obs} = Q_{obs} - Q_{ENN}^{(1)} $, it follows that $ Q_{NN} = Q^{(2)}_{NN} + Q^{(1)}_{ENN}$, and $(Q^{(2)}_{NN} - Q^{(2)}_{obs}) = (Q_{NN} - Q_{obs})$.

Then, we can further reduce the leakage by creating new maps with almost no E-mode contribution:
\begin{align}
    Q^{(3)}_{obs} = & (Q_{obs} - Q_{ENN}^{(1)}-Q_{ENN}^{(2)}) \\
    U^{(3)}_{obs} = & (U_{obs} - U_{ENN}^{(1)}-U_{ENN}^{(2)}). 
\label{QU2}
\end{align}
Finally, if we use the maps $Q^{(3)}_{obs}$ and $U^{(3)}_{obs}$, the contribution now is primarily from the B-mode rather than the E-mode. The outputs of the neural network with these maps is $Q_{NN}^{(3)}$ and $U_{NN}^{(3)}$, which allow us to obtain the Wiener Filter of the B-mode. Besides, it is possible to use the results from the iterations to further improve the Wiener Filter of the E-mode, obtaining the E-mode contribution from $Q_{NN} = Q_{ENN}^{(1)} + Q_{ENN}^{(2)} + Q_{NN}^{(3)}$ and $U_{NN} = U_{ENN}^{(1)} + U_{ENN}^{(2)} + U_{NN}^{(3)}$. This iterative process can be repeated as many times as needed to achieve better performance. 

\begin{figure}
\centering
\includegraphics[width=0.7\columnwidth]{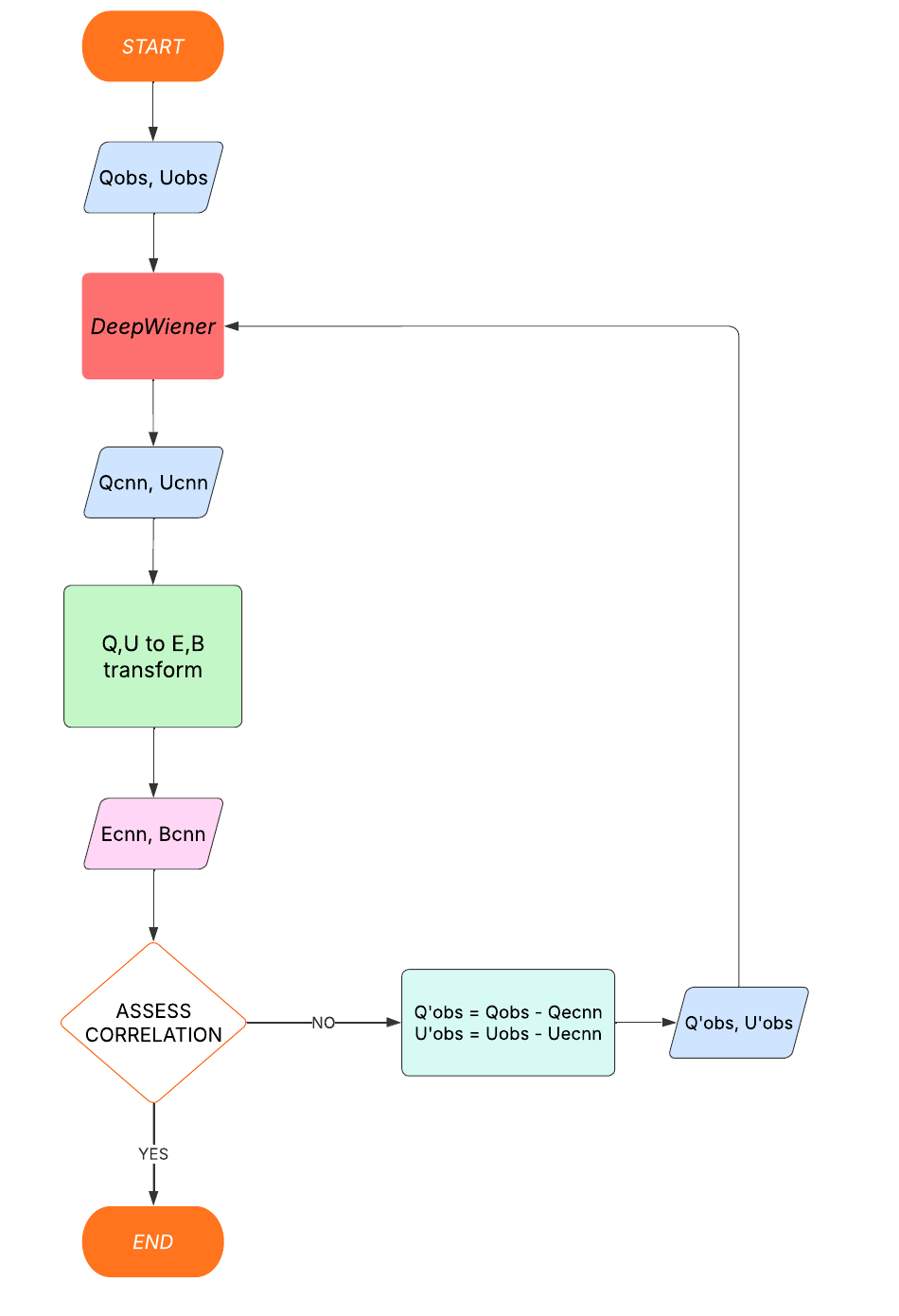}
\caption{The flow of the iterative process is structured as follows. The dataset composed of $Q$ and $U$ maps are presented in blue, while the $E$ and $B$ maps in pink. There are three main processes in data processing: in red, the DeepWiener neural network processes the $Q$ and $U$ maps; in green, these maps are transformed into $E$ and $B$ modes; and in turquoise, a new dataset is generated with the E-mode contribution removed. Finally, the calculation of the cross-correlation coefficient determines whether the process ends or needs to be repeated with the new data, ensuring that the correlation of the B-mode is satisfactory before concluding the iteration.}
\label{flow}
\end{figure}

A flowchart of the described iterative process is presented in Figure~\ref{flow}. It begins with the observed $Q_{obs}$ and $U_{obs}$ maps, which are used for training and evaluating the proposed neural network, DeepWiener, as explained in Section~\ref{sec:deepwiener}. The outputs of DeepWiener $Q_{CNN}$, $U_{CNN}$, are then transformed into $E_{CNN}$ and $B_{CNN}$ maps. 

To assess the quality of these maps, we compute the cross-correlation coefficient between the WF from the CNN and the WF from the CG. If the cross-correlation, particularly for the B-mode, is close to 1, the process terminates. Otherwise, a new dataset is generated with the E-mode contribution removed, as described in equation~\ref{QU1}. The process then restarts with these new maps and continues iteratively until a satisfactory cross-correlation coefficient is achieved.

On the other hand, when we obtain observed maps $Q$ and $U$ with minimal E-mode contribution, we can assume that the polarization information comes predominantly from the B-mode contribution, therefore the transformation from spin-2 quantities $Q$ and $U$ to scalar quantities will yield an E-mode map nearly zero. Thus, we can treat the problem similarly to the temperature case where the neural network was trained with a temperature map as input; in this case it would be a B-mode map. Then, instead of training the neural network with $Q_{obs}^{(3)}$ and $U_{obs}^{(3)}$ using the loss function $J_{Q,U}$, we can train using the B-mode map, $B_{obs}^{(3)}$, that is derived from the observed maps $Q_{obs}^{(3)}$ and $U_{obs}^{(3)}$ considering the following loss function: 
\begin{equation}
    J_B = \sum_{i}^{N_{pix}} \frac{\left(B_{NN} - B_{obs}^{(3)}\right)^{2}}{\sigma_{i}^{*2}} +  \sum_{\ell} \frac{B_{lNN}B_{lNN}^*}{C_{l}^B}.
\label{polB}
\end{equation}
It is important to note that the noise variance map $\sigma^{2}$ for $Q$ and $U$ maps is not the same as the noise variance map $\sigma^{*2}$ for $E$ and $B$ maps. We can approximate $\sigma^{*2}$ by computing the variance in each pixel of the difference $B_{obs}-B_{sky}$ over thousands of simulations, as described in the section~\ref{sec:datasets}, where $B_{obs}$ is derived from $Q_{obs}$ and $U_{obs}$. The output of the neural network will directly provide the WF of the B-mode,  and in this case, fewer iterations are required compared to those needed when using the loss function $J_{Q,U}$, as explained in section~\ref{sec:results}.

%\textcolor{red}{This expression is an approximation where the B-mode map is not contaminated by \textit{ambiguous} modes and the $\sigma^{*2}$ is estimated through simulations. It is not derived from~\eqref{pol}.}
This expression is not derived from~\eqref{pol} but rather is an approximation where the B-mode map is not contaminated by \textit{ambiguous} modes and the $\sigma^{*2}$ is estimated through simulations.

\subsection{Power spectrum estimation}

In CMB data analysis, it is of interest to compress the information in the field into a summary statistic, such as the power spectrum. After reconstructing the underlying signal $\hat{\textbf{s}}$ using a neural network that efficiently simulates the optimal WF method, we aim to obtain the power spectrum amplitudes (or band-powers $\bi{\Theta}$ if binned) and their covariance matrix using the optimal quadratic estimator. For that purpose, we follow the approach outlined in~\citep{2017JCAP...12..009S, 2019JCAP...10..035H}. 

To find the most probable set of band-power $\bi{\Theta}$ given the measurements \textbf{d}, we need to maximize the likelihood function L(\textbf{d}|$\bi{\Theta}$), which is proportional to the posterior P($\bi{\Theta}$|\textbf{d}) assuming a flat prior on $\bi{\Theta}$. Since the modes are Gaussian distributed, the likelihood function can be expressed as: 
\begin{equation}
    L(\textbf{d}|\bi{\Theta}) = (2\pi)^{-N/2}det(C)^{-1/2}exp(-\frac{1}{2}\textbf{d}^{\dag}C^{-1}\textbf{d}),
\label{likelihood}
\end{equation}
where the covariance matrix of the data can be expressed as a linear sum over the band-powers, with their response matrix $P_{\ell}$ for each mode:
%\textcolor{red}{[$P_{\ell}$ es una nueva base que depende de $\Pi_{\ell}$, el operador proyeccion que te pasa de S a los bandpowers. Ver paper Horowitz y Seljak]}): 
%entiendo que C no es diagonal en ninguna base porque es la suma de S y N (cada una diagonal en bases distintas)
\begin{equation}
    C = \textbf{R}\textbf{S}\textbf{R}^{\dag} + \textbf{N} = \sum_{\ell} \Theta_{\ell}P_{\ell} + \textbf{N}.
\end{equation}

The maximum of the likelihood~\eqref{likelihood} is the optimal quadratic estimator, obtained using Newton's method: 
\begin{equation}
    F_{\ell\ell'}\Theta_{\ell} = \frac{1}{2} \textbf{d}^{\dag}C^{-1}P_{\ell'}C^{-1}\textbf{d}-\frac{1}{2}Tr[C^{-1}P_{\ell'}C^{-1}N], 
\label{optimal}
\end{equation}
where $F^{-1}$ is the local estimate of the covariance matrix of the band-power parameters, 
\begin{equation}
    F^{-1} = \langle \bi{\Theta \Theta}^{\dag} \rangle - \langle \bi{\Theta} \rangle \langle \bi{\Theta}^{\dag} \rangle.
\end{equation}
In practice, this analytical expression for the likelihood is inefficient, as its evaluation requires inversion or computing the determinant of a large, non-sparse matrix. Therefore, in our approach, we first worked with the variables \textbf{s}, solving the WF with the neural network, and then marginalized over these variables to obtain the likelihood of the parameters.

To do so, we need to introduce a derivative matrix $\bi{\Pi}_\ell$ around some fiducial power spectrum $\textbf{S}^{\rm fid}$, defined as: %
\begin{equation}
\left[\frac{\partial \textbf{S} }{\partial \Theta_\ell}\right]_{\textbf{S}^{\rm fid}}=\bi{\Pi}_\ell.
\label{pi}
\end{equation}

Then, the true covariance can be written as: 
\begin{equation}
\textbf{S}=\textbf{S}^{\rm fid}+\sum_\ell\Delta \Theta_\ell\bi{\Pi}_\ell.
\end{equation}

Note that for linear dependence of $\textbf{S}$ on $\bi{\Theta}$ we can use 
\begin{equation}
\bi{\Pi}_\ell=\frac{\textbf{S}_{\rm fid} }{\Theta_\ell},  
\label{pilin}
\end{equation}
i.e. $\langle s_{k_\ell}s_{k_\ell}^* \rangle=\Theta_\ell\Pi_l(k_\ell)$, where, in this case, the derivative matrix take us from $\Theta_\ell$ (the power spectrum value representative over a bin) to $\textbf{S}$, which is the power spectrum.  

Then, we expand the log-likelihood in terms of $\bi{\Theta}$ to  quadratic order around some fiducial values $\bi{\Theta}_{\rm fid}$ (marginalized over \textbf{s}):
\begin{equation}
\ln L(\bi{\Theta}_{\rm fid}+\Delta \bi{\Theta})=\ln L(\bi{\Theta}_{\rm fid})+\sum_\ell \left[\frac{\partial \ln L(\bi{\Theta})}{\partial \Theta_\ell} \right]_{\bi{\Theta}_{\rm fid}}\Delta \Theta_\ell+ \frac{1}{2} \sum_{\ell \ell'}\left[\frac{\partial^2 \ln L(\bi{\Theta})}{\partial \Theta_\ell \partial \Theta_{\ell'}}\right]_{\bi{\Theta}_{\rm fid}}\Delta \Theta_\ell \Delta \Theta_{\ell'},
\label{llik}
\end{equation}
where the last term of equation \eqref{llik} defines the curvature matrix as the second derivatives of log-likelihood with respect to the parameters.

We define: 
\begin{equation}
E_\ell(\textbf{S}_{\rm fid},\hat{\textbf{s}})=\frac{1}{2}\hat{\textbf{s}}^{\dag}\textbf{S}_{\rm fid}^{-1}\bi{\Pi}_\ell\textbf{S}_{\rm fid}^{-1}\hat{\textbf{s}} = \frac{1}{2}\sum_{k_\ell}\frac{\hat{s}_{k_\ell}^2 }{\Theta_{\rm{fid},\ell} S_{{\rm fid},k_\ell}},
\label{el}
\end{equation}
where the sum over $k_\ell$ accounts for all the modes that contribute to the band-power $\Theta_\ell$. Then, the first derivative of the likelihood becomes: 
\begin{equation}
\frac{\partial \ln L(\bi{\Theta}) }{\partial \Theta_\ell}=E_\ell - b_\ell. 
\label{like_deriv}
\end{equation}

The maximum likelihood solution for $\bi{\hat{\Theta}}$ is obtained setting the equation above equal to zero leading to the noise bias term equal to:
\begin{equation}
b_\ell= E_\ell(\bi{\Theta}_{\rm fid},\hat{\textbf{s}}_{s+n}),
\label{blmv}
\end{equation}
where $\hat{\textbf{s}}_{s+n}$ is the wiener-filtered map obtained from data that contains signal and noise ($s+n$), where the signal map is a realization of the \textit{fiducial} power spectrum.

Finally, the peak of the likelihood function is obtained by setting the derivative of \eqref{llik} with respect to $\Delta \bi{\Theta}$ equal to zero, and using the equation \eqref{like_deriv}, leading to: 
\begin{equation}
    (\textbf{F}\Delta\bi{\Theta})_{\ell} = E_\ell - b_\ell,
    \label{corr}
\end{equation}
where the signal map appearing in $E_\ell$ is a realization of the \textit{true} power spectrum. This approach provides an unbiased estimator, as squaring the raw power spectrum requires subtracting the noise bias term. Additionally, the Fisher matrix accounts for both the covariance matrix and the band-power mixing, given that we are working with incomplete sky coverage in the signal estimation.

The bias term and the Fisher matrix are calculated with simulations of the \textit{fiducial} power spectrum, and applying the WF to each of them. The estimation of the \textit{true} power spectrum will be $\bi{\Theta}_{fid} + \Delta\bi{\hat{\Theta}}$. 

%\textcolor{red}{[Claudia: Me parece que la sección del power spectrum es la más difícil de leer (y de digerir!)]}

In practice, the measurements consist of $Q$ and $U$ polarization maps, which are filtered through the neural networks models to obtain the maps $Q_{WF}$ and $U_{WF}$, and subsequently $E_{WF}$ and $B_{WF}$. Our goal is to estimate the auto-spectra of both the E-mode and B-mode, as these are rotationally invariant and thus more suitable for capturing the polarization information. Therefore, we need to compute equation~\eqref{corr} for each field: 
\begin{equation}
    (\textbf{F}^{E}\Delta\bi{\Theta}^{E})_{\ell} = E_\ell^{E} - b_\ell^{E},
    \label{corrE}
\end{equation}
\begin{equation}
    (\textbf{F}^{B}\Delta\bi{\Theta}^{B})_{\ell} = E_\ell^{B} - b_\ell^{B},
    \label{corrB}
\end{equation}
and estimate the \textit{true} power spectrum for each of them: $\bi{\Theta}^{E}_{fid} + \Delta\bi{\hat{\Theta}}^{E}$, $\bi{\Theta}^{B}_{fid} + \Delta\bi{\hat{\Theta}}^{B}$.

In section~\ref{sec:namaster} we compare the uncertainties in the power spectrum estimation calculated with the procedure described above, with the power spectrum estimation using the \textsc{Namaster} library~\citep{2019MNRAS.484.4127A}. The last approach approximates the covariance matrix in equation~\eqref{optimal} for its diagonal, assuming uncorrelated data. Some approximations and analytical expressions in flat-sky limit will be discussed and compared with our method in section~\ref{sec:namaster}.

\section{DeepWiener}
\label{sec:deepwiener}

\begin{figure}
\centering
\includegraphics[width=1.\columnwidth]{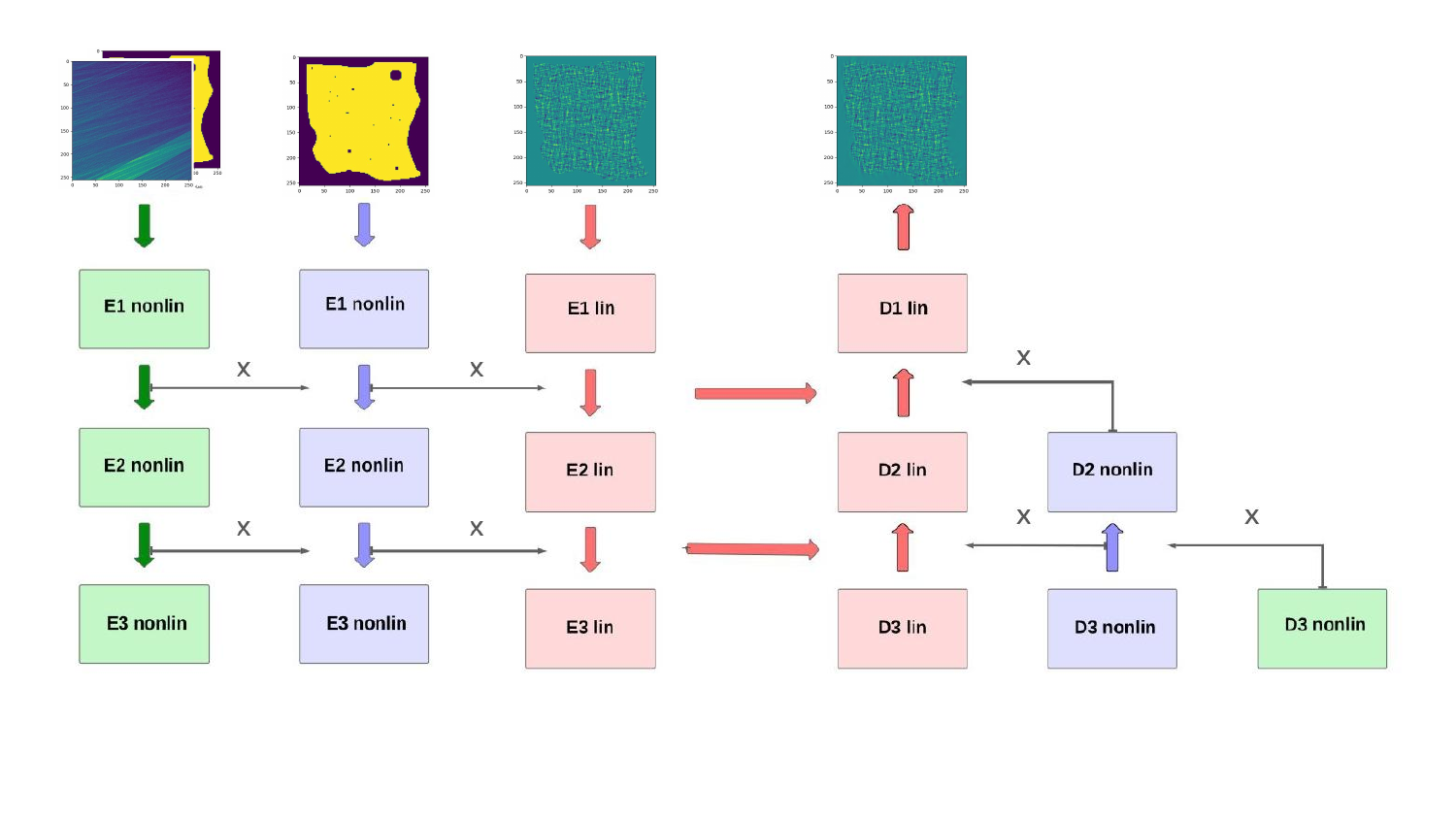}
\caption{Scheme of DeepWiener architecture: only three encoders are presented for visualization.  The black solid line represents the multiplication between channels, and the red arrow connecting the encoder part to the decoder part represents the skip connections in the linear channel. These skip connections are also present in the other channels although not being drawn here. 
The linear channel can be applied to the $Q$ and $U$ maps or the $B$ mode map. The third channel is two-dimensional and contains both the variance map and the mask.
}
\label{esquema}
\end{figure}

DeepWiener is the neural network developed to perform the WF on polarization data with inhomogeneous noise. The architecture presented in Figure~\ref{esquema} is the same as the one used for temperature maps with inhomogeneous noise in~\citep{2024JCAP...04..041C}, originally based on WienerNet~\citep{2019arXiv190505846M}. 

The architecture is an autoencoder Convolutional Neural Network (CNN), consisting of an encoder made up of convolutional layers and a decoder composed of transposed convolutional layers. This type of neural network is commonly used in image analysis for processing visual data. During the encoder phase, the network extracts key features from the input data while reducing its dimensions to form a latent space representation. Conversely, the decoder phase increases the layer dimensions to reconstruct an output image of the same size as the input.

This network follows a UNet architecture~\citep{2015arXiv150504597R}, but with two additional non-linear channels that operate on the mask and the variance map, while the linear channel is exclusively for the CMB maps. The outputs from the non-linear channel are multiplied by those from the map channel. This architecture is motivated by the need to perform linear operations on the CMB maps, as the WF is a linear filter applied to the data, as shown in equation~\eqref{eq:eqWF}. 

The network receives as inputs the observed polarization maps, $Q_{obs}$ and $U_{obs}$, with the corresponding mask applied and inhomogeneous noise added, assuming that the loss function $J_{Q,U}$ (equation~\eqref{pol}) is used. The outputs will be the filtered maps, $Q_{WF}$ and $U_{WF}$. If the loss function $J_{B}$ (equation~\eqref{polB}) is employed, the input consists of the observed B-mode map, which results from subtracting the E-mode contribution to the $Q_{obs}$ and $U_{obs}$ maps. In that case, the output is directly the filtered B-mode map, $B_{WF}$. 

The key distinction from the procedure applied to temperature maps in~\citep{2024JCAP...04..041C} lies in addressing the E-to-B leakage present in polarization maps. To handle this, we created a new dataset with the $E_{WF}$ contribution removed and retrained the neural network multiple times, each time removing the obtained $E$ contribution, forcing the CNN to focus on the B-mode. This approach was motivated by the need to mitigate the \textit{ambiguous} B-modes by enhancing the reconstruction of the E-modes that are present in the maps.

We have made the DeepWiener implementation and the power spectrum procedure publicly available in a dedicated GitHub repository\footnote{\url{https://github.com/Belencostanza/DeepWiener}}.

\section{Datasets}
\label{sec:datasets}

%\subsection{Maps}
In this paper, we have built the training set considering polarization maps with inhomogeneous noise applied, as an extension of previous work with inhomogeneous noise but only on temperature maps. We adopt the flat-sky approximation and simulate maps of $20^\circ \times 20^\circ$ and $256 \times 256$ pixels. The angular resolution is determined by the size of the map and the number of pixels considered. 

\begin{figure}
\centering
\includegraphics[width=1.\columnwidth]{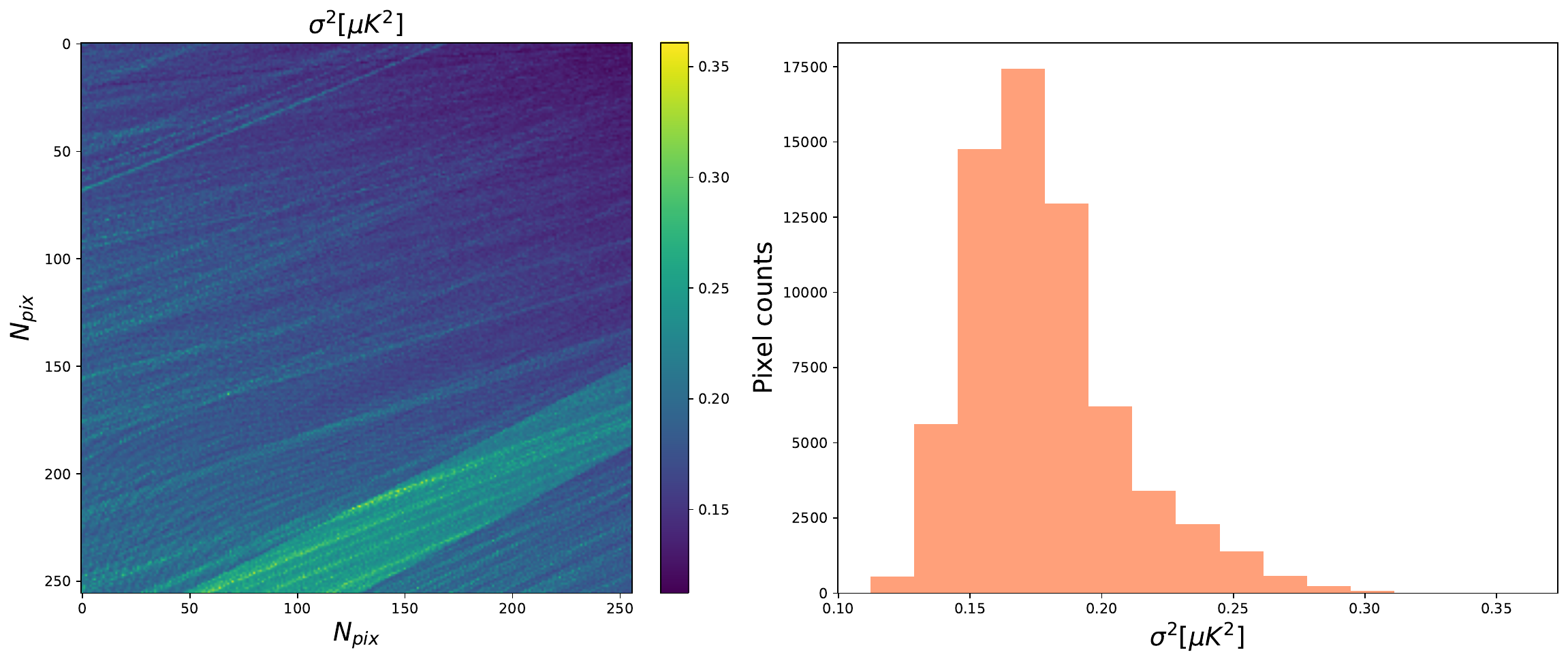}
\caption{Left panel: Variance map extracted from Planck. Right panel: Histogram of the variance map showing the distribution of pixel counts
across different noise variances.}
\label{inho_hist}
\end{figure}

\begin{figure}
\centering
\includegraphics[width=1.\columnwidth]{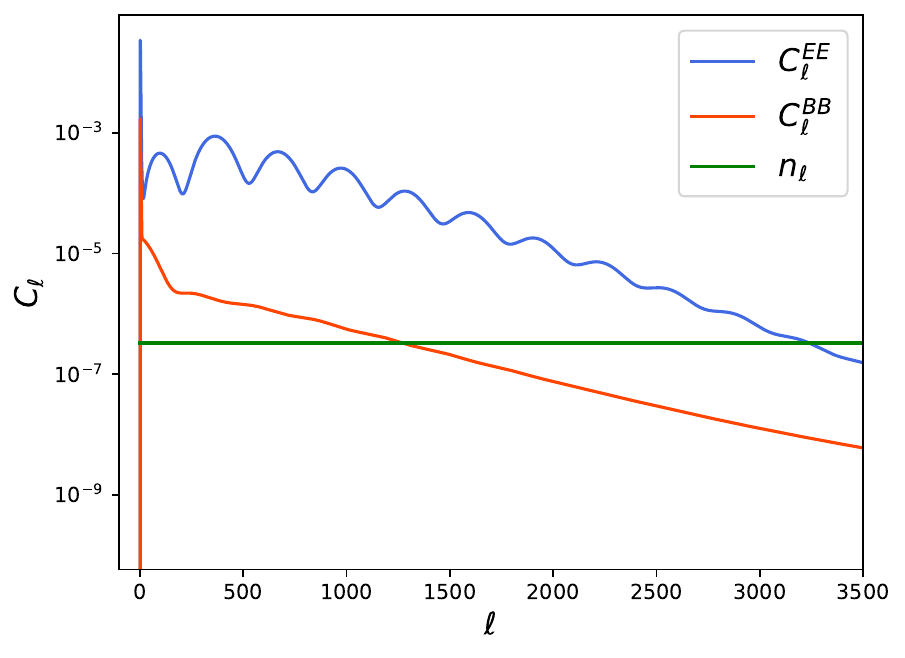}
\caption{Power spectrum of the E-mode and B-mode signal and average noise level of the variance map extracted from Planck. }
\label{espectros}
\end{figure}

%We extracted a map variance from Planck noise maps~\citep{2020A&A...641A...1P} to simulate the inhomogeneous noise to assess a realistic case, as illustrated in the left panel of Figure~\ref{inho_hist}, while the distribution of variance levels among pixels is presented in the histogram of the right panel. 
We extracted a map variance (hit maps) from Planck noise maps~\citep{2020A&A...641A...1P} to simulate the inhomogeneous noise to asses a realistic case. Since the maps from Planck are in the curved sky, it was necessary to perform a gnomonic projection, using the Healpix~\citep{2005ApJ...622..759G, Zonca2019} library\footnote{\url{https://healpix.sourceforge.io/index.php}}, providing the number of pixels and the desired resolution. The resulting map is illustrated in the left panel of Figure~\ref{inho_hist}, while the distribution of variance levels among pixels is presented in the histogram of the right panel.
Figure~\ref{espectros} presents the \textit{fiducial} power spectrum of the E-mode and B-mode signal obtained using CAMB~\citep{2000ApJ...538..473L} with the cosmological parameters from the best fit of Planck data (2014) from Table 5 in~\citep{planck2013}. In addition, it is presented the average noise level of the variance map, which cuts the B-mode power spectrum on scale $\ell \approx 1260$. A rescaling was applied to the variance map values to obtain a mean noise level that effectively cuts the B-mode power spectrum, as the noise levels in Planck measurements are quite high for polarization.

%\textcolor{red}{[CLAUDIA] Creo que acá faltaría una oración diciendo que primero entrenas con la $J_{Q,U}$ para ir obteniendo el modo E y restandoselo a los mapas observados, hasta que la contribución de E es despreciable luego de varias iteraciones, y que los mapas de Q y U solo contengan contribucion de B. Ver sugerencia en magenta:}

\begin{figure}
\centering
\includegraphics[width=1.\columnwidth]{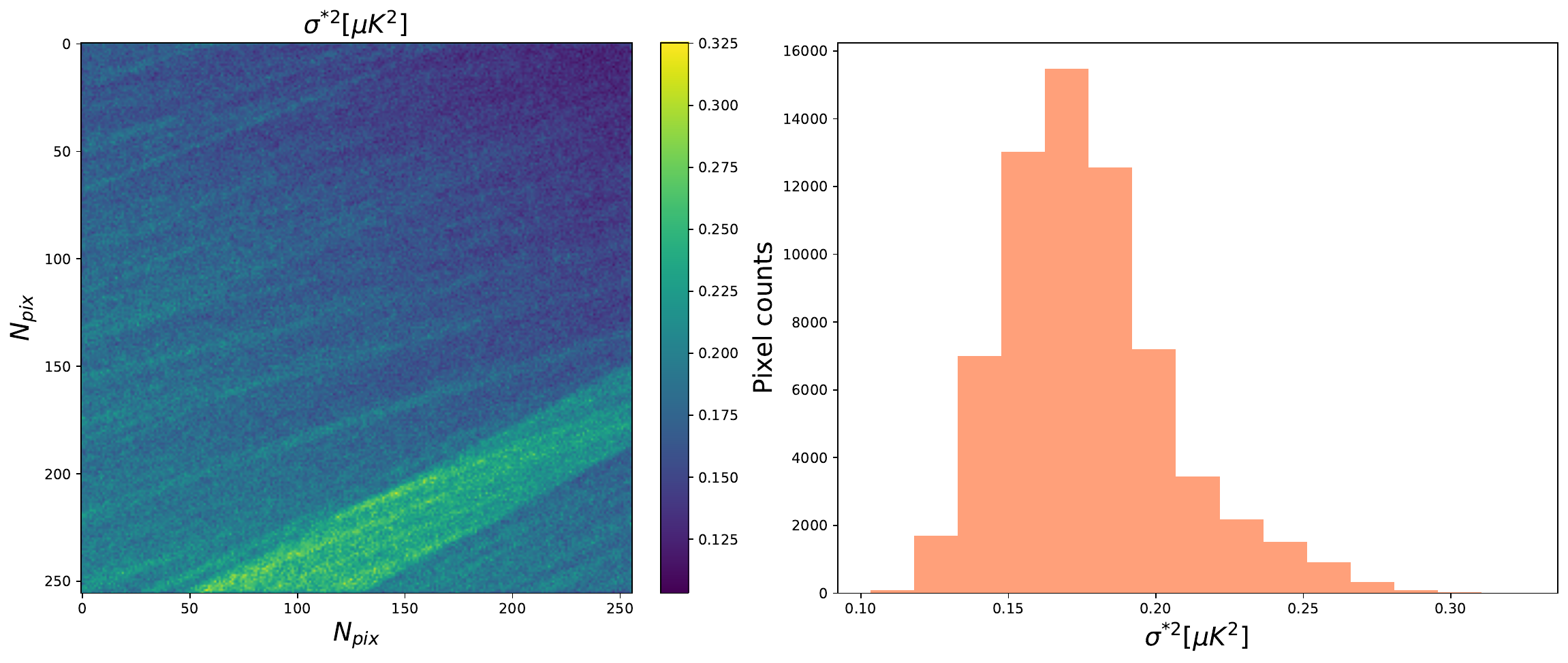}
\caption{Left panel: Variance map for the B-mode map calculated from the difference $B_{obs}-B_{sky}$. Right panel: Histogram of the variance map for the B-mode map showing the distribution of pixel counts
across different noise variances.}
\label{inho_b}
\end{figure}

Initially, the neural network is trained with the observed $Q$ and $U$ maps to iteratively reconstruct the $E$-mode component, which is then subtracted from the observed maps. This process is repeated over several iterations until the $E$-mode contribution becomes negligible, leaving $Q$ and $U$ maps that primarily contain the $B$-mode signal. At this stage, it becomes possible to train the neural network using the $B$-mode map derived from these polarization maps with the E-mode effectively removed.
%
%\sout{When a dataset with minimal E-mode contribution is created, it becomes possible to train the neural network using the B-mode map derived from the polarization maps with the E-mode subtracted. }
%
Under this assumption, DeepWiener is trained with the loss function $J_B$, equation~\eqref{polB}, where $\sigma^{*2}$ represents the variance map for the E-mode and B-mode maps, differing from the variance maps of the $Q$ and $U$ maps. Figure~\ref{inho_b} shows the variance map for each pixel in the B-mode map and the corresponding histogram, computed from the difference of $B_{obs}-B_{sky}$. The intensity values of $\sigma^{*2}$ differ from those of $\sigma^{2}$, but the inhomogeneity pattern remains consistent. 

\begin{figure}
\centering
\includegraphics[width=1.\columnwidth]{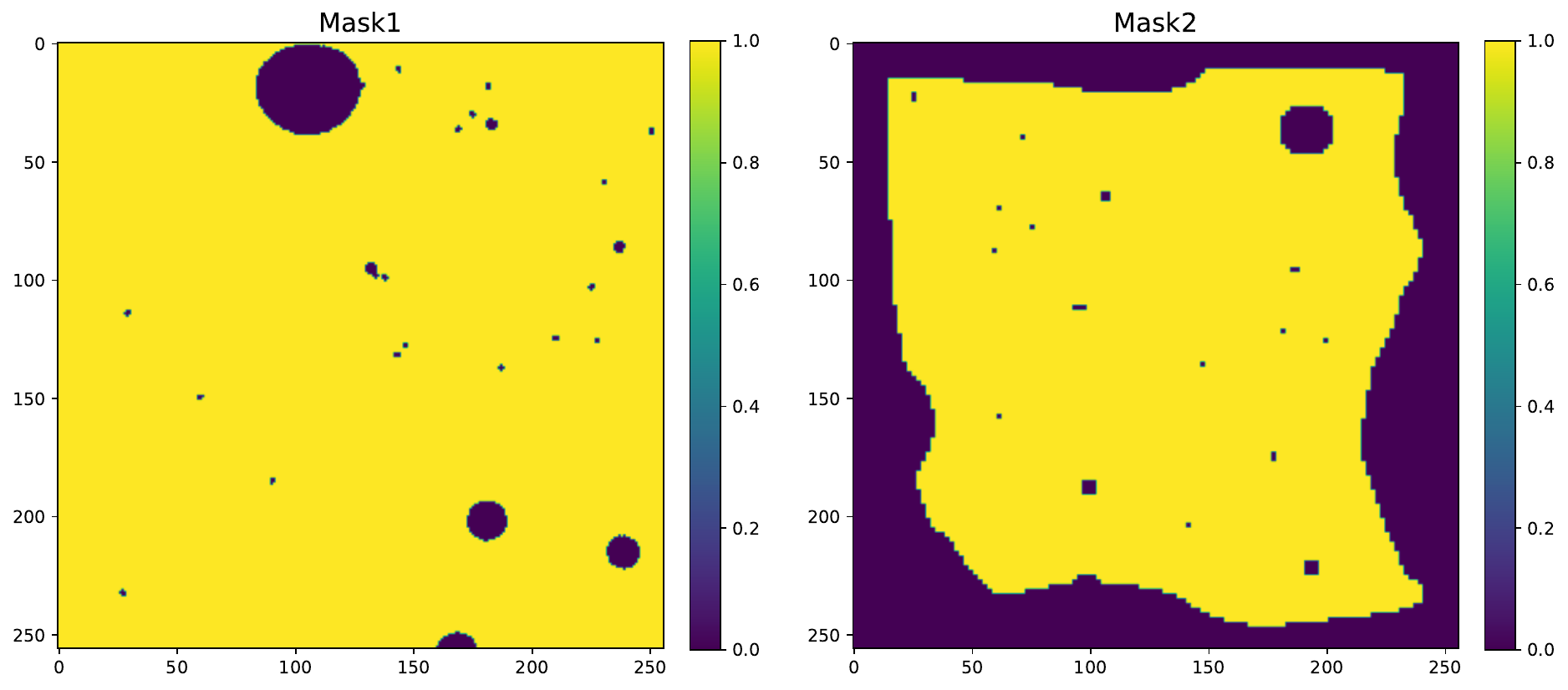}
\caption{Left panel: mask with point sources of different sizes. Right panel: mask with tiny point sources and edges.}
\label{mascaras}
\end{figure}

Since the E-to-B leakage arises from the binary mask, we study the performance of the neural network in estimating the Wiener Filter of E and B considering two masks with different complexities and fractions of the sky. The left panel of Figure~\ref{mascaras} shows a mask with only point sources of different sizes, while the right panel presents a mask with point sources and edges, covering a significant portion of the sky.

After the training, the models are used to estimate the power spectrum of maps with a different \textit{true} angular power spectrum, obtained by slightly altering the cosmological parameters from their fiducial values. These maps, created using the \textit{true} power spectrum, were not part of the training data, as the goal is to estimate the power spectrum of an unknown underlying signal. Each neural network used in this work was trained with
the Apollo GPU nodes of the Institute for Advanced Studies, where each node contains 8 GPUs NVIDIA A100.

\section{Results}
\label{sec:results}

\subsection{Comparison with PCG results}

We trained DeepWiener using \textsc{Tensorflow} 2 and \textsc{Python} 3 on maps with $256 \times 256$ pixels experimenting with different hyperparameter values to ensure the minimization of the loss function. We obtained satisfactory results with a learning rate $lr = 3.36 \times 10^{-5}$ and a number of filters between 16 and 32 in each layer. %\textcolor{magenta}{We don't discard the possibility to find even better results expanding the set of hyperparameters tested, at the expense of more computing time for training. Algunos hiperparametros probé con optuna pero no una cantidad significativa ya que es una red que tarda bastante y es pesada, los que tomé fueron los mejores. No descartó la posibilidad que puedan haber incluso mejores por explorar. A su vez en algún lado habría que poner las capas y layers para 256x256}.

We constructed polarization datasets for training and validation, using the two masks shown in Figure~\ref{mascaras}, to evaluate the performance of the neural network under different mask complexities and sky coverage. In the initial training, we successfully reconstructed the E-mode WF but performed poorly on the B-mode due to the dominant E-mode contribution in the polarization maps. To address this, we retrained the network using a new dataset with the E-mode contribution from the first training removed. To mitigate the E-to-B leakage, we repeated this process multiple times, with each iteration removing the E-mode contribution obtained in the previous one. As anticipated, the more complex mask (Mask2) required more iterations (five), while the simpler mask (Mask1) required less iterations (four), using the loss function $J_{Q,U}$ defined in equation~\eqref{pol}. Refer to Appendix~\ref{apx3} for detailed information about iterations results.

\begin{figure}
\centering
\includegraphics[width=1.\columnwidth]{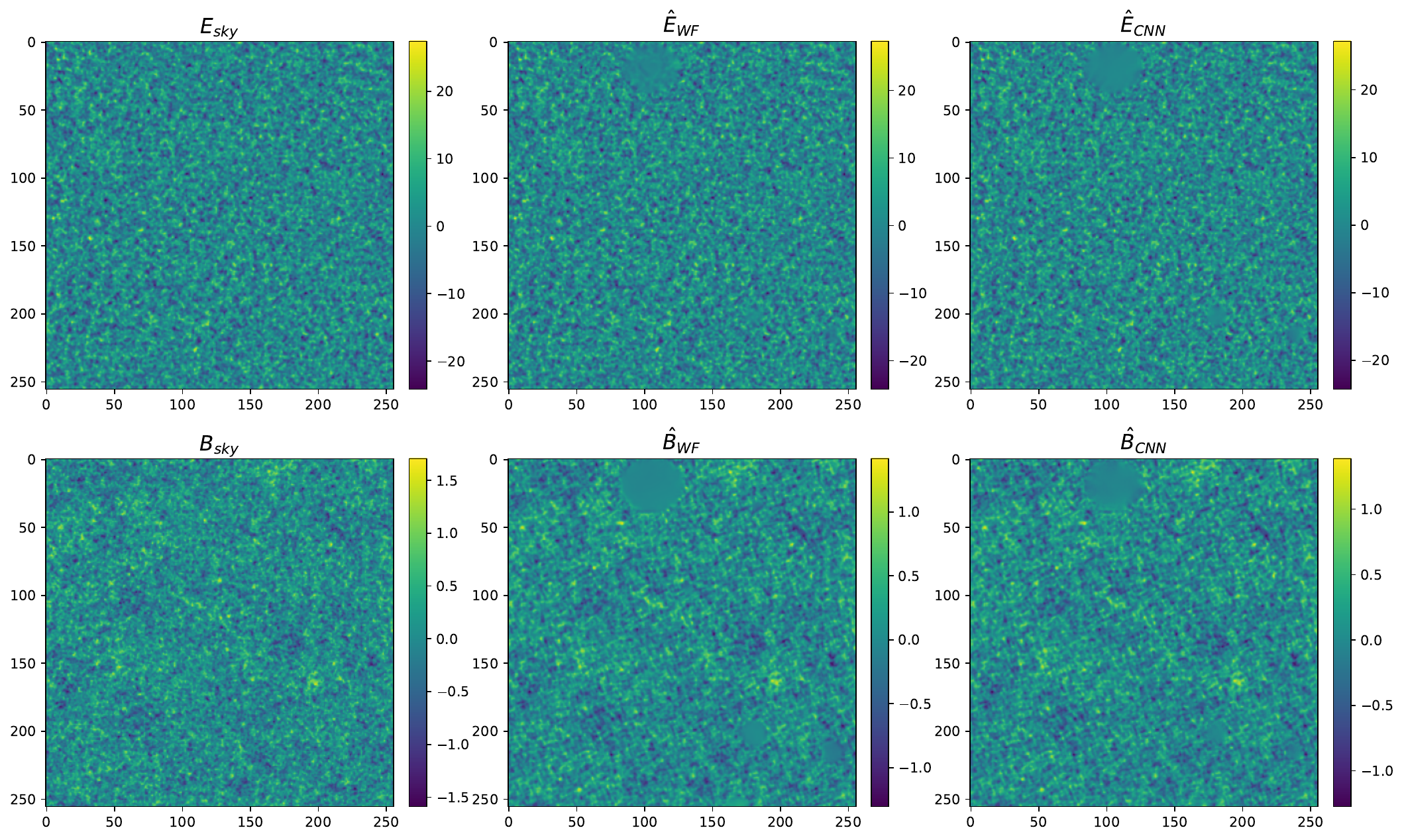}
\caption{On the left panel it is presented an E-mode map signal and a B-mode map signal, simulated using the \textit{fiducial} power spectrum. On the center it is presented the WF reconstruction with the PCG method for both modes with Mask1 applied, while on the right panel it is presented the reconstruction for both modes using DeepWiener and the loss function $J_{Q,U}$, equation~\eqref{pol}.}%which is qualitatively similar with the reconstruction using DeepWiener and the loss function~\ref{pol}, as it can be seen on the E-mode and B-mode maps of the right panel.}
\label{eb_mask1}
\end{figure}

Figure~\ref{eb_mask1} at the top compares a map of the E-mode signal, generated from the \textit{fiducial} power spectrum, with the WF map computed with the PCG method and with the trained DeepWiener using the Mask1. On the bottom, it is the comparison of the B-mode signal with the corresponding WF reconstruction using the PCG method and the neural network. We can visually check that the outputs of DeepWiener and the procedure followed to address the E-to-B leakage on polarization maps are accurate to reconstruct the signal in unmasked pixels while reconstructing large-scale modes even in the masked region near the edges. These results were trained with the loss function $J_{Q,U}$ from equation~\eqref{pol}.

\begin{figure}
\centering
\includegraphics[width=1.\columnwidth]{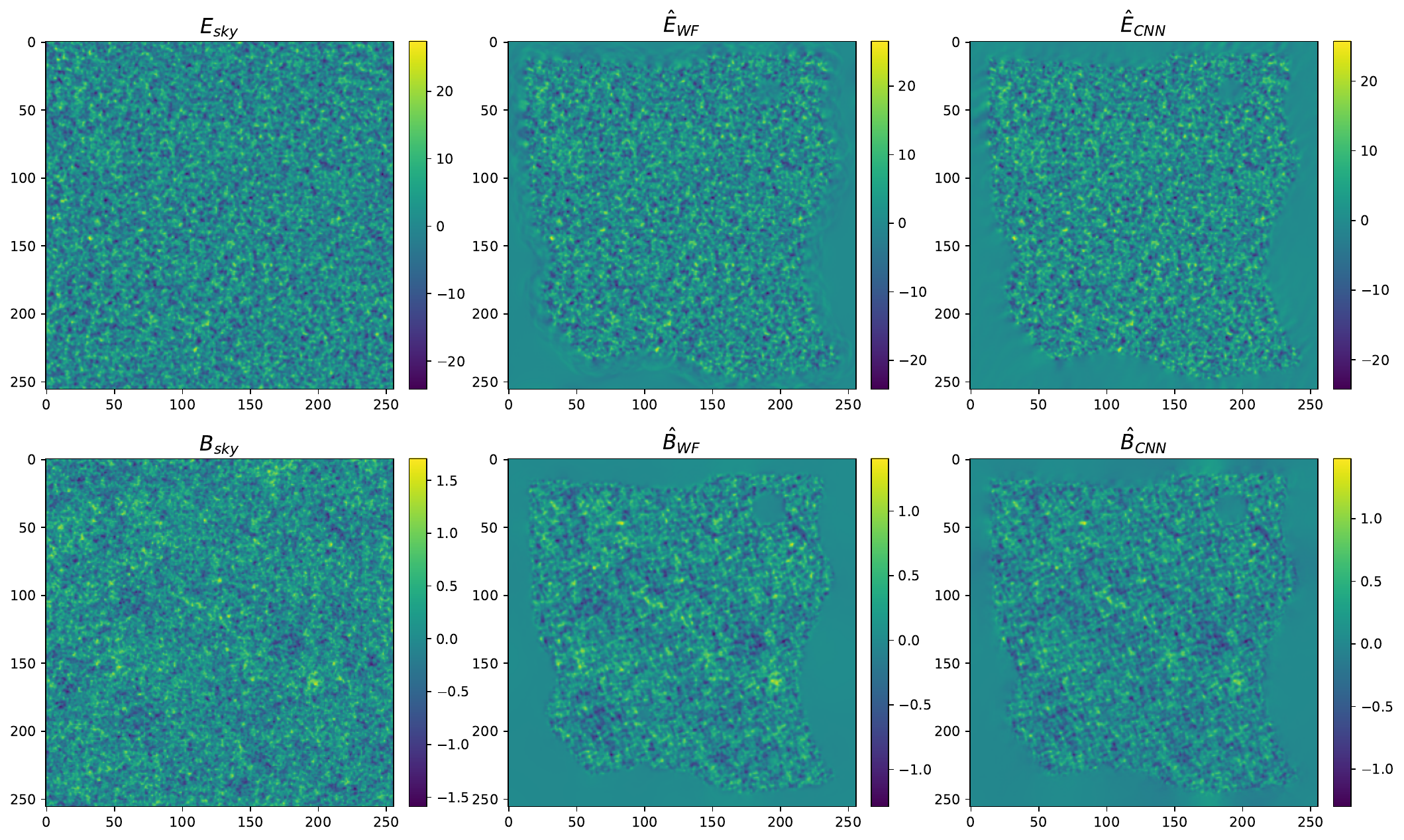}

\caption{The left panel shows the E-mode and B-mode map signals, obtained from the \textit{fiducial} power spectrum. The center panel presents the WF reconstruction using the PCG method for both fields with Mask2 applied. In the right panel, the reconstruction for both fields is shown using DeepWiener with the loss function $J_{Q,U}$ from equation~\eqref{pol}.}
%\caption{On the left panel it is presented an E-mode map signal and a B-mode map signal, obtained from the \textit{fiducial} power spectrum. On the center it is presented the WF reconstruction with the PCG method for both modes with Mask2 applied, while on the right panel it is presented the reconstruction for both modes using DeepWiener and the loss function $J_{Q,U}$, equation~\eqref{pol}.}% which is qualitatively similar with the reconstruction using DeepWiener and the loss function~\ref{pol}, as it can be seen on the E-mode and B-mode maps of the right panel.}
\label{eb_mask2}
\end{figure}

Figure~\ref{eb_mask2} presents the same comparison as in Figure~\ref{eb_mask1}, but with a more complex mask (Mask2) applied to the data. Qualitatively, it is evident that the E-mode reconstruction using PCG captures slightly more information near the edges compared to the E-mode results obtained with DeepWiener. Additional training epochs may improve this, but the cross-correlation results indicate that the performance is already satisfactory, as it will be shown later. For the B-mode reconstruction, an additional iteration was required compared to Mask1 to achieve the results shown in the lower right panel, which visually are very similar to the PCG results.
%Figure~\ref{eb_mask1_jb} illustrates a comparison between signal reconstruction using DeepWiener trained with the loss function~\ref{pol} (middle panel), and DeepWiener trained with the loss function~\ref{polB} (right panel). Visually, both reconstructions are similar, indicating that both loss functions effectively capture the necessary B-mode signal information. However, the main difference is that with the loss function~\ref{polB}, only 3 training iterations were required (the first two using $Q$ and $U$ maps, and the last one using the B-mode map), compared to 4 iterations with the loss function~\ref{pol}. This suggests that the neural network more easily performs the Wiener filtering when the B-mode map is directly provided.
%
%\begin{figure}
%\centering
%\includegraphics[width=1.\columnwidth]{ewf_bwf_mask1_JB.pdf}
%%\caption{The left panel and center panel are the same maps as in Figure~\ref{eb_mask1}, but now compared with the reconstruction using DeepWiener and the loss function~\ref{polB}, presented on the right panel.}
%\label{eb_mask1_jb}
%\end{figure}

The agreement in the WF simulated with DeepWiener and computed with PCG can be quantified using the cross-correlation coefficient $r_{\ell}$ as a function of the multipole $\ell$. This coefficient measures how well the two methods align across different angular scales and is defined as: 
\begin{equation}
    r_{\ell} = \frac{\langle a_{CNN}(\ell) a^*_{WF}(\ell)\rangle}{\sqrt{\langle a_{CNN}(\ell) a^*_{CNN}(\ell)\rangle \langle a_{WF}(\ell) a^*_{WF}(\ell)\rangle}},
\end{equation}
where $a_{CNN}$ refers to the discrete Fourier coefficients of either the E-mode or B-mode, depending on the output of the neural network. If the loss function $J_{Q,U}$ in equation~\eqref{pol} is used, $a_{CNN}$ will be the E-mode or B-mode Fourier coefficients obtained from the $Q$ and $U$ maps filtered by the neural network. Alternatively, when the loss function $J_{B}$ in equation~\eqref{polB} is used, $a_{CNN}$ will represent the B-mode Fourier coefficients derived from the B-mode output map directly. Then, the cross-correlation coefficient $r_{\ell}$ is averaged over the test set (10 maps). This averaging provides a more robust measure of how well the two WF methods agree across different realizations.

\begin{figure}
    \centering
    \includegraphics[width=1.\columnwidth]{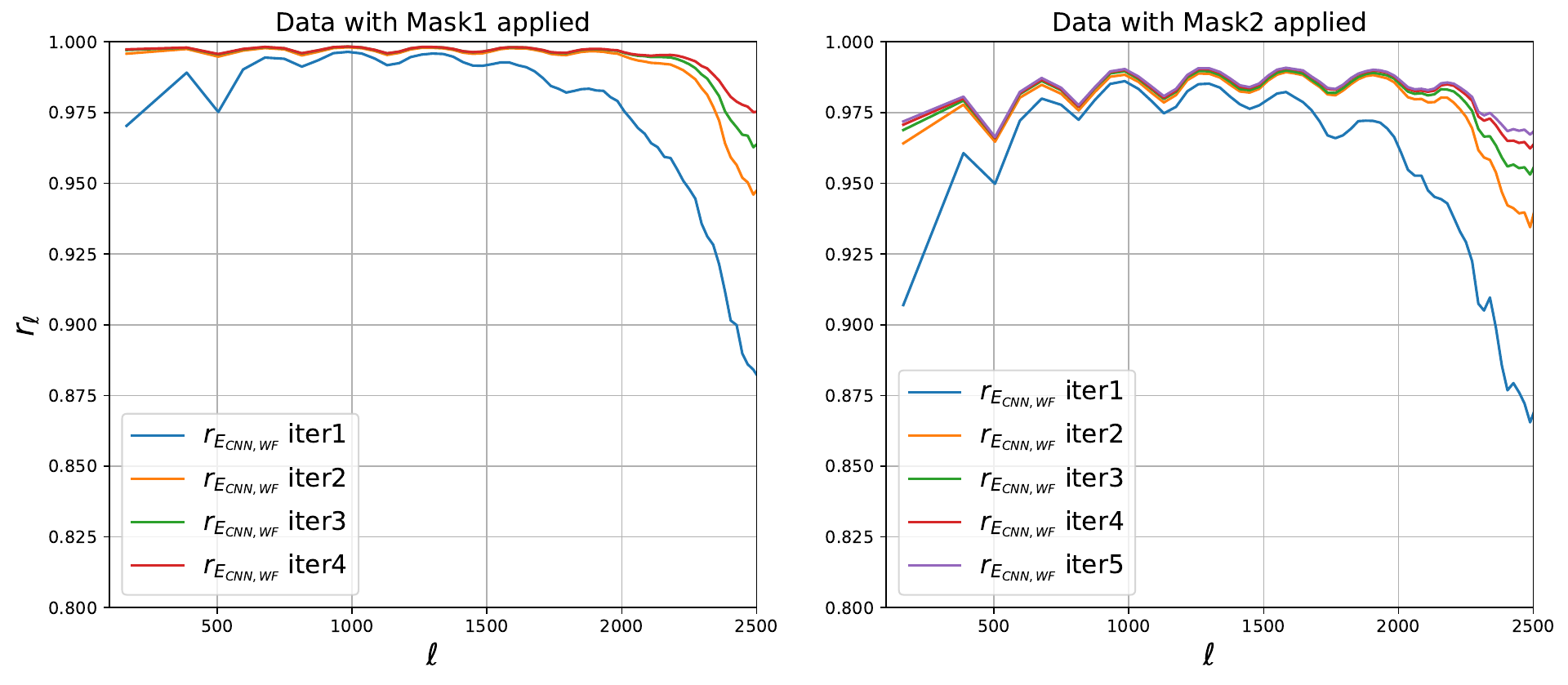}
    \caption{Cross-correlation coefficient for the E-mode map between the DeepWiener models and the WF with PCG, for each iteration. The left panel presents the results for data with Mask1 applied, while the right panel shows the results for Mask2.}% In both cases, the improvement with increasing iterations is evident, with Mask2 requiring one additional iteration compared to Mask1.}
    \label{rle}
\end{figure}

The cross-correlation coefficient for the E-mode is presented in Figure~\ref{rle}, for data with Mask1 (left panel) and Mask2 (right panel) applied. With each iteration, the E-mode reconstruction improves as corrections are applied to the results from the previous iteration. Mask2 presents a more extended and complex structure compared to Mask1, resulting in a more pronounced E-to-B leakage. Training additional epochs or performing more iterations is expected to be necessary to effectively mitigate leakage and achieve accurate signal reconstruction. More intermediate results are presented in Appendix~\ref{apx3}. 

\begin{figure}
    \centering
    \includegraphics[width=1.\columnwidth]{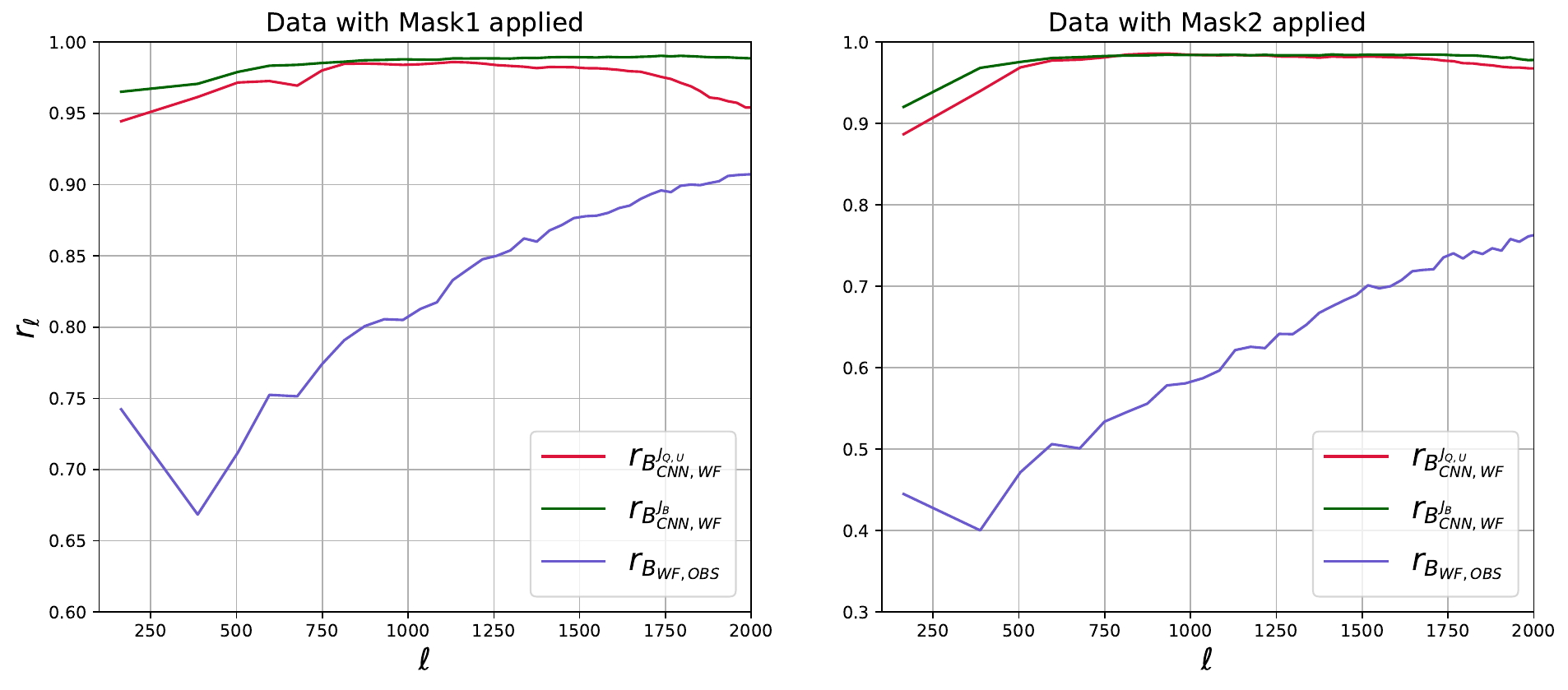}
    \caption{Cross correlation coefficient for the B-mode map between the DeepWiener models and the WF with PCG. The left panel presents the results for data with Mask1 applied, while the right panel shows the results for Mask2. In both cases, the final iteration result is shown where the E-mode contribution has been nearly removed. The red and green curves represent the cross-correlation using DeepWiener trained with the loss functions $J_{Q,U}$~\eqref{pol} and $J_{B}$~\eqref{polB}, respectively. Additionally, the cross-correlation coefficient between the observed B-mode map and the B-mode WF is also included for comparison.}
    \label{rlb}
\end{figure}

The E-mode reconstruction with the neural network shows a high correlation with the PCG method, which is expected since there is minimal noise contribution on the scales of interest, as shown in Figure~\ref{espectros}. Figure~\ref{rlb} shows the cross-correlation coefficient for the B-mode, with Mask1 applied in the left panel and Mask2 in the right panel. In both cases, the noise becomes dominant at scales beyond $\ell \approx 1260$.

It is clear that the cross-correlation between the observed B-mode map (transformed from $Q_{obs}$ and $U_{obs}$, with significant E-to-B leakage) and the WF computed using PCG highlights the improvement achieved by applying the DeepWiener models, especially on larger scales. Notably, the B-mode map reconstructed using DeepWiener trained with the loss function $J_{B}$ (green curve) shows a higher correlation compared to the model trained with $J_{Q,U}$ (red curve). This indicates that the network performs better when a B-mode map is used as input instead of $Q$ and $U$, leading to more accurate signal reconstruction. Besides, the number of iterations for the B-mode reconstruction using the loss function $J_{B}$ is one less than using the loss function $J_{Q,U}$ (3 iterations for data with Mask1 and 4 iterations for data with Mask2).

In the next section, we will explore whether the performance of DeepWiener models with different loss functions and mask has any impact on the estimation of the power spectrum. 

On Appendix~\ref{apx4} it is presented the computational time required to perform the WF using DeepWiener with several models, compared with the computational time using the PCG method.

%On the other hand, Mask2 presents a more complex structure compared to Mask1, resulting in a more pronounced E-to-B leakage. It is expected the necessity to train more epochs or perform more iterations to effectively mitigate the leakage and achieve accurate signal reconstruction. 

\subsection{Implementation of the power spectrum}
\label{sec:spectra}

%Después de tener las redes con el algoritmo definitivo y bien optimizadas, vamos a hacer el cálculo de power spectrum, como hicimos en el paper anterior. No creo que acá tengamos dificultades, porque una vez que tengamos los mapas E y B, son campos escalares frente a rotaciones, como el caso de la temperatura. El power spectrum se calcula de la misma manera que para la temperatura. 

%Entonces resumimos la metodologia presentada en el paper anterior, en la sección de Metodolodogía, y acá van los resultados. 

For the estimation of the power spectrum, it is necessary to assess the noise bias and the Fisher matrix for both E-mode and B-mode, as presented in section~\ref{sec:methodology}. This evaluation takes place after the estimation of the underlying field $\hat{s}$, for which the selection of a \textit{fiducial} angular power spectrum is required. Then, we computed the noise bias and Fisher matrix through simulations of the \textit{fiducial} power spectrum. 

We started generating a Gaussian realization of the signal in Fourier space, for the E-mode and B-mode, denoted as $\textbf{s}^{E}_{s}$ and $\textbf{s}^{B}_{s}$ respectively:  
\begin{align}
    \langle |\textbf{s}^{E}_{s}|^2 \rangle = & \textbf{S}^{E}_{fid} \\
    \langle |\textbf{s}^{E}_{s}|^2 \rangle = & \textbf{S}^{B}_{fid}. 
\label{ebfid}
\end{align}

Then, we transformed properly to $Q$ and $U$ data since the received observations are not in the ($E$,$B$) basis. We generated random noise realizations using the variance map presented in Figure~\ref{inho_hist}:
\begin{align}
    \textbf{d}^{Q}_{s+n} = & \textbf{d}^{Q}_{s} + \textbf{d}^{Q}_{n} \\
    \textbf{d}^{U}_{s+n} = & \textbf{d}^{U}_{s} + \textbf{d}^{U}_{n}, 
\end{align}
where we have considered that the noise realization is the same for $Q$ and $U$, therefore $\textbf{d}^{Q}_{n}$ = $\textbf{d}^{U}_{n}$.

We applied the DeepWiener models on these dataset and transformed to obtain $\hat{\textbf{s}}^{E}_{s+n}$ and $\hat{\textbf{s}}^{B}_{s+n}$. We have calculated the noise bias term for both modes, with the expression~\eqref{blmv}, averaged over several realizations, that enables to obtain an unbiased estimator of the power spectrum. 

In the presence of a mask, the Fisher matrix accounts for the band-power mixing and can be interpreted as the response of the band-power $\ell$ to another band-power $\ell'$. A small perturbation is introduced to a Gaussian realization of the \textit{fiducial} power spectrum at a specific band-power $\ell'$, leading to the following equation:
\begin{equation}
    F_{\ell \ell'}\Delta \Theta_{\ell'} = E_{\ell}(\Theta_{fid}, \hat{\textbf{s}}_{\ell',s+n})-E_{\ell}(\Theta_{fid}, \hat{\textbf{s}}_{s+n}), 
\label{eq:fisher}
\end{equation}
that should be calculated for both E and B-modes.

After computing the noise bias and the Fisher matrix through simulations, averaged over several realizations, we estimated a new unknown \textit{true} power spectrum, which was never part of the neural network training, using the expressions~\eqref{corrE} and~\eqref{corrB}. 

\begin{figure}
\centering
\includegraphics[width=1.\columnwidth]{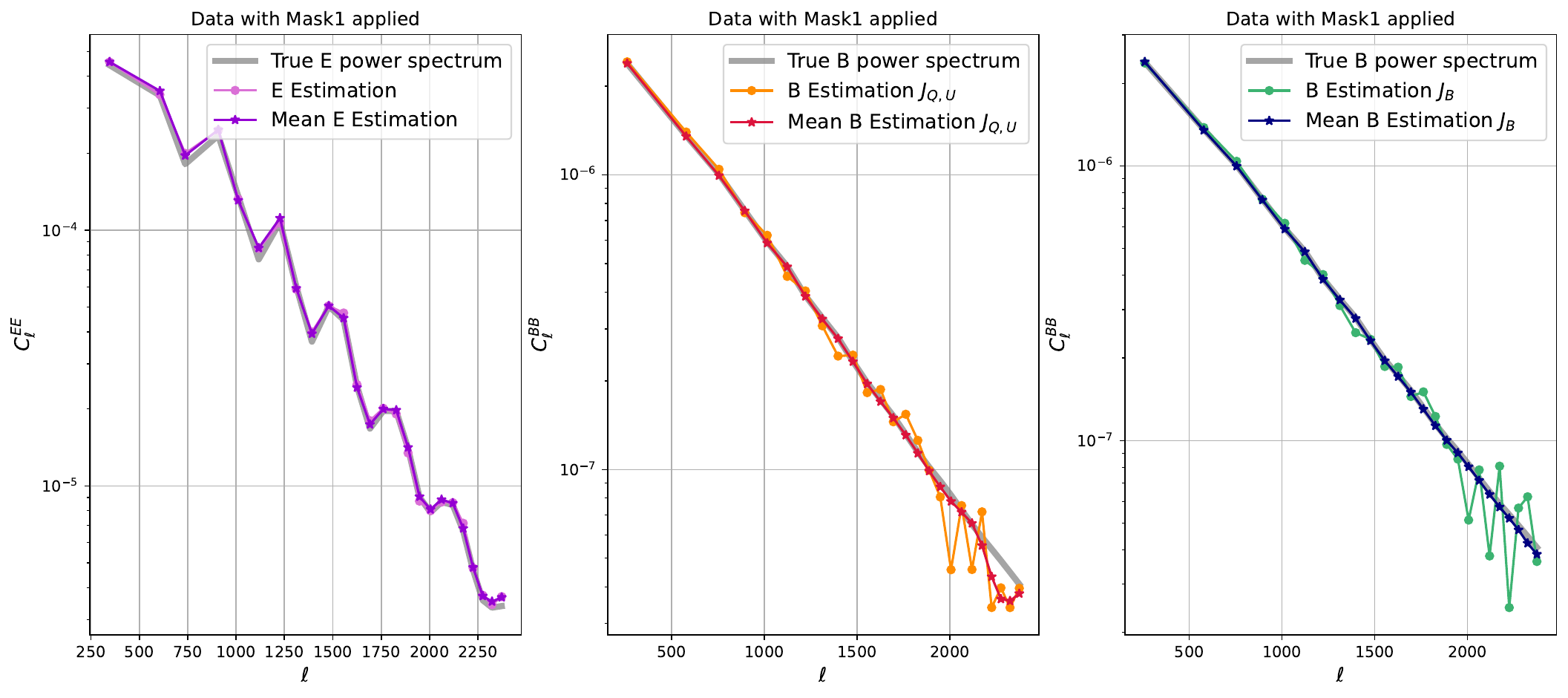}
\caption{Left panel: estimation of the true E power spectrum for one map and the average estimation over 100 maps. Middle panel: estimation of the true B power spectrum for one map and the average over 100 maps, using DeepWiener models trained with $J_{Q,U}$~\eqref{pol}. Right panel: estimation of the true B power spectrum for one map and the average over 100 maps using DeepWiener models trained with $J_{B}$~\eqref{polB}.}
\label{cl_mask1}
\end{figure}

\begin{figure}
\centering
\includegraphics[width=1.\columnwidth]{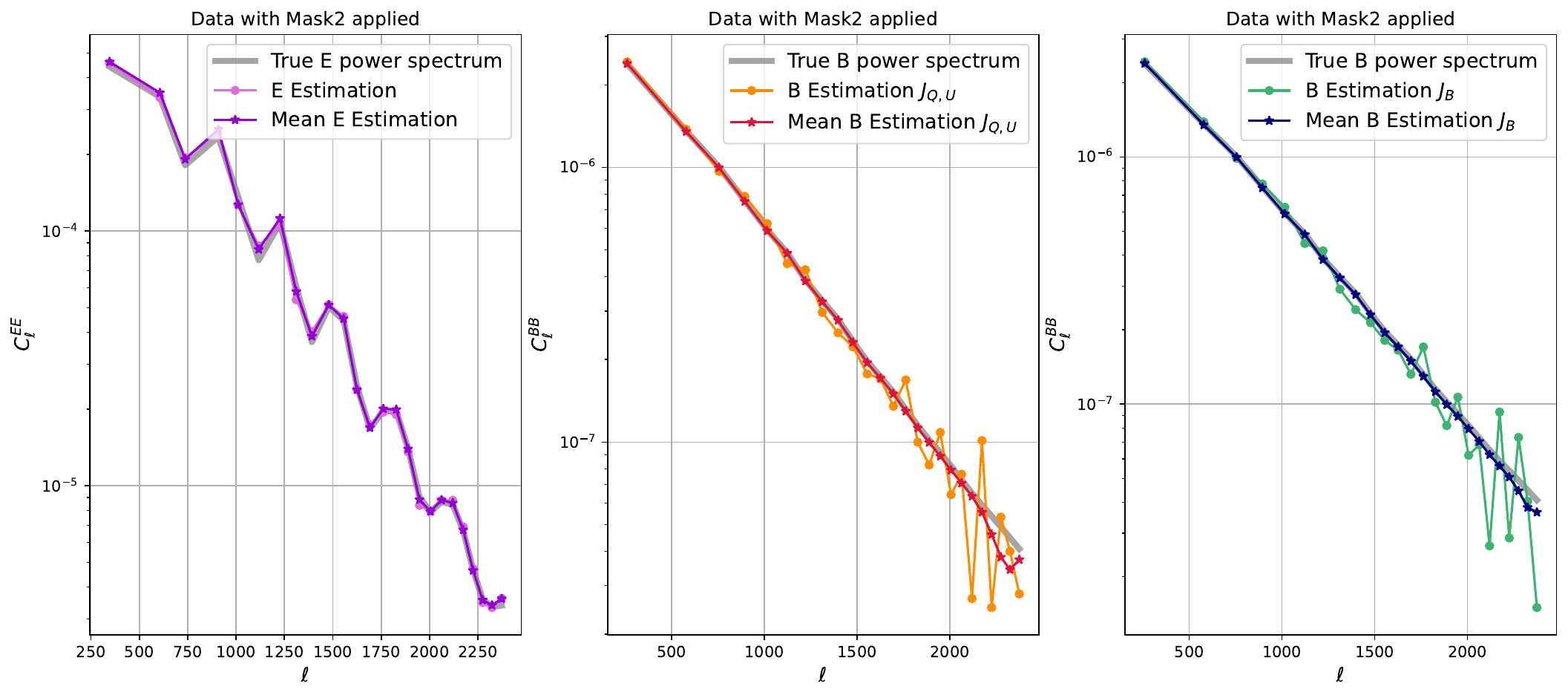}
\caption{Left panel: estimation of the true E power spectrum for one map and the average estimation over 100 maps. Middle panel: estimation of the true B power spectrum for one map and the average over 100 maps, using DeepWiener models trained with $J_{Q,U}$~\eqref{pol}. Right panel: estimation of the true B power spectrum for one map and the average over 100 maps using DeepWiener models trained with $J_{B}$~\eqref{polB}.}
\label{cl_mask2}
\end{figure}

Figure~\ref{cl_mask1} and~\ref{cl_mask2} show the power spectrum estimation for both E-modes and B-modes, for a single map with Mask1 and Mask2 applied, and averaged over 100 maps. In the left panel, the E-mode case is presented, where there is no noise contribution at any scale, leading to an estimation that matches the \textit{true} power spectrum. The middle and right panels display the B-mode case using models trained with $J_{Q,U}$ and $J_{B}$, respectively. On average, the B-mode estimation remains unbiased across all scales, although the estimation for individual maps becomes noisier at scales far beyond $\ell \approx 1260$, where the noise dominates. For now on, we will focus our analysis for the B-mode power estimation.

\begin{figure}
\centering
\includegraphics[width=1.\columnwidth]{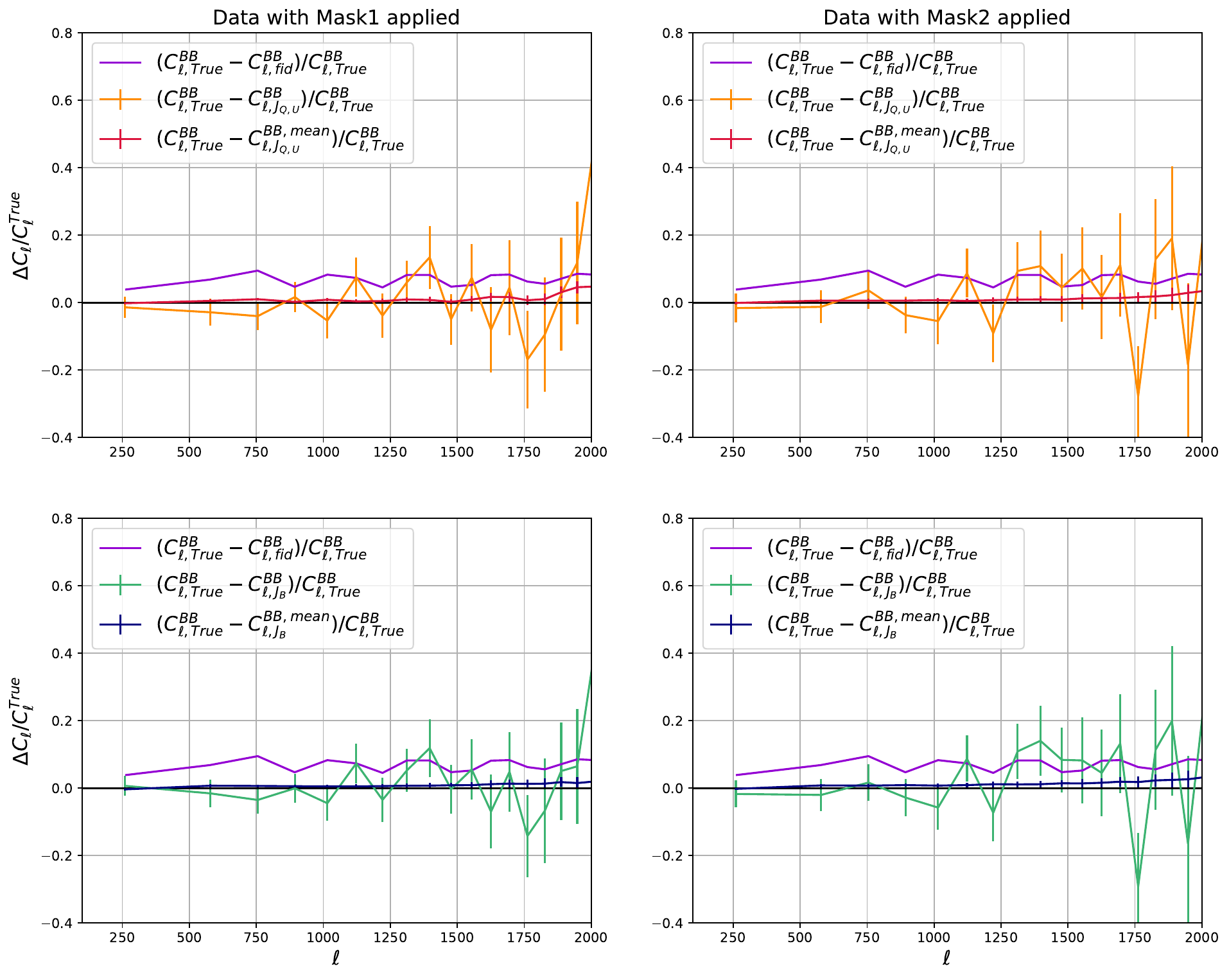}
\caption{Relative difference between the true B power spectrum and the estimation of the power spectrum for models trained with $J_{Q,U}$~\eqref{pol} (top panels) and $J_{B}$~\eqref{polB} (bottom panels), for data with Mask1 (left panels) and Mask2 (right panels) applied. It is presented, in each panel, the relative difference for the average estimation over 100 maps with the corresponding error bar, and the relative difference for the estimation of a single map.} %where it is notable that the estimation is around zero and unbiased. Also, it is presented the relative difference for the estimation of a single map.}%, where the error bars are larger for one measurement and increase when the noise becomes predominant respect to the signal.}
\label{deltacl_mask1_mask2}
\end{figure}

To compare the ability of the DeepWiener models, trained with different loss functions, to estimate the \textit{true} B-mode power spectrum it is more clear to analyze the relative difference between the estimated B-mode power spectrum and the \textit{true} power spectrum, as shown on Figure~\ref{deltacl_mask1_mask2}. The \textit{fiducial} power spectrum serves as a starting point from which the unknown \textit{true} power spectrum is estimated by applying a correction $\Delta \bi{\Theta}$, that is computed through the outlined procedure. Consequently, the purple line represents the relative difference between the \textit{true} and \textit{fiducial} power spectra, reflecting a bias by construction. 

The left panels in Figure~\ref{deltacl_mask1_mask2} display the relative difference between the estimated B-mode power spectrum and the \textit{true} power spectrum for data with Mask1 applied, while the right panels show the same comparison for data with Mask2 applied. For all cases it is presented a single measurement and the average over 100 maps, along with the corresponding error bars. It is noticeable that the estimations are unbiased for both loss functions, as the values remain centered around zero, specially for data with Mask1 applied. However, the error bars grow larger at scales beyond $\ell \approx 1260$, where the noise level becomes dominant compared to the signal, and the signal-to-noise ratio is quite smaller than one. This demonstrates that both loss functions are effective for estimating $B_{WF}$ and its corresponding power spectrum. 

\begin{figure}
\centering
\includegraphics[width=1.\columnwidth]{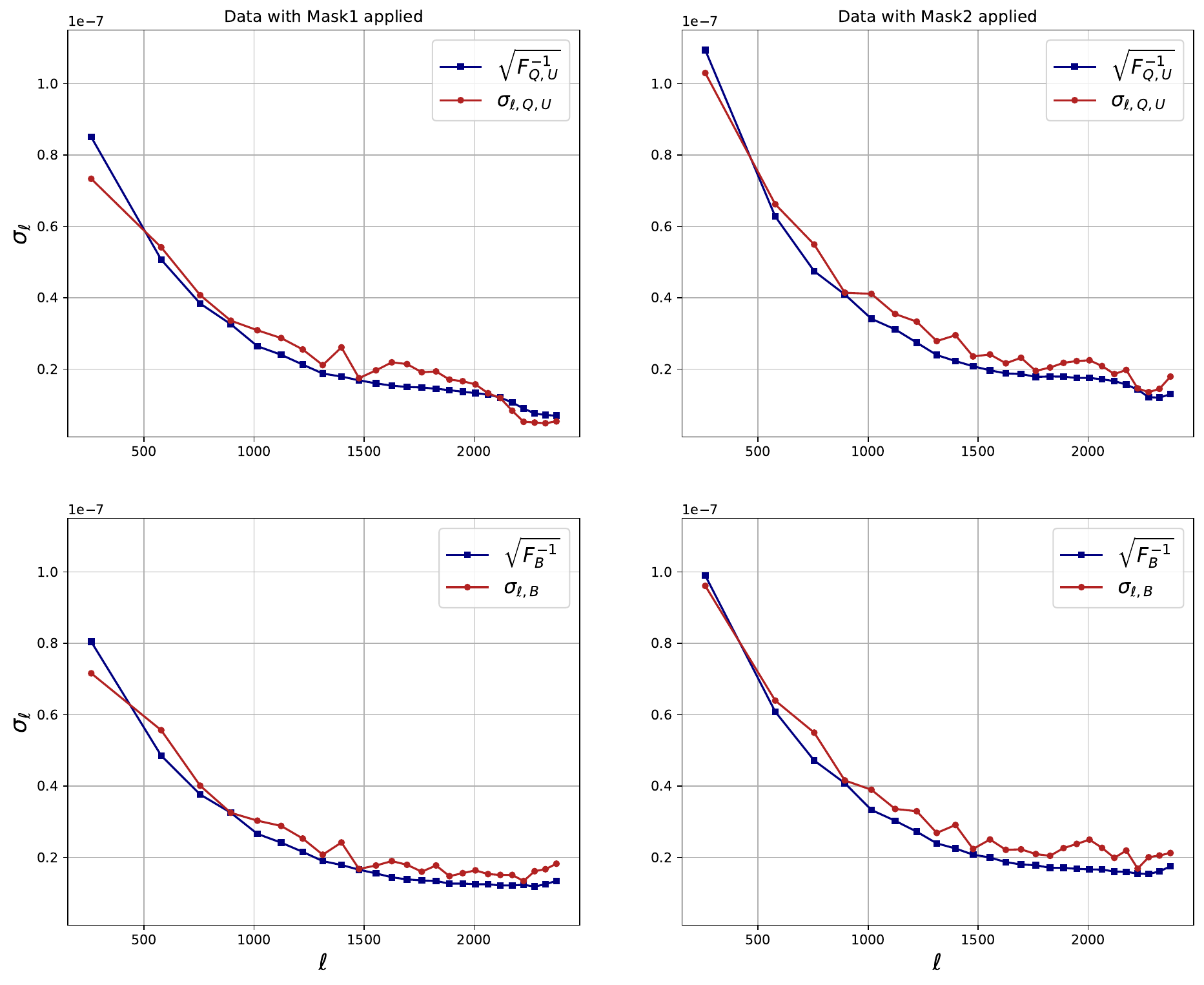}
\caption{Square root of the diagonal part of the inverse Fisher matrix and the error of the estimation of the \textit{true} power spectrum, for models trained with $J_{Q,U}$~\eqref{pol} (top panels) and $J_{B}$~\eqref{polB} (bottom panels).}
\label{desvcl_mask1_mask2}
\end{figure}

It is worthwhile to compare the errors of the estimated power spectrum for a single map with the square root of the inverse of the Fisher matrix, $F^{-1}$. The inverse Fisher matrix can be interpreted as an estimate of the covariance matrix of the elements $\Theta_{\ell}$, assuming the modes are Gaussian distributed. Therefore, in Figure~\ref{desvcl_mask1_mask2}, the square root of the inverse Fisher matrix, calculated using equation~\eqref{eq:fisher}, is presented for models trained with $J_{Q,U}$ (top panels) and $J_{B}$ (bottoms panels). The left panels show the comparison for data with Mask1 applied, while on the right panel it is presented for data with Mask2 applied. As expected, it closely matches the error in the estimation of the \textit{true} power spectrum. A small difference can be observed since the Fisher matrix corresponds to the error of the \textit{fiducial} power spectrum, while we are estimating an unknown power spectrum referred to as the \textit{true} power spectrum. 

\begin{figure}
\centering
\includegraphics[width=1.\columnwidth]{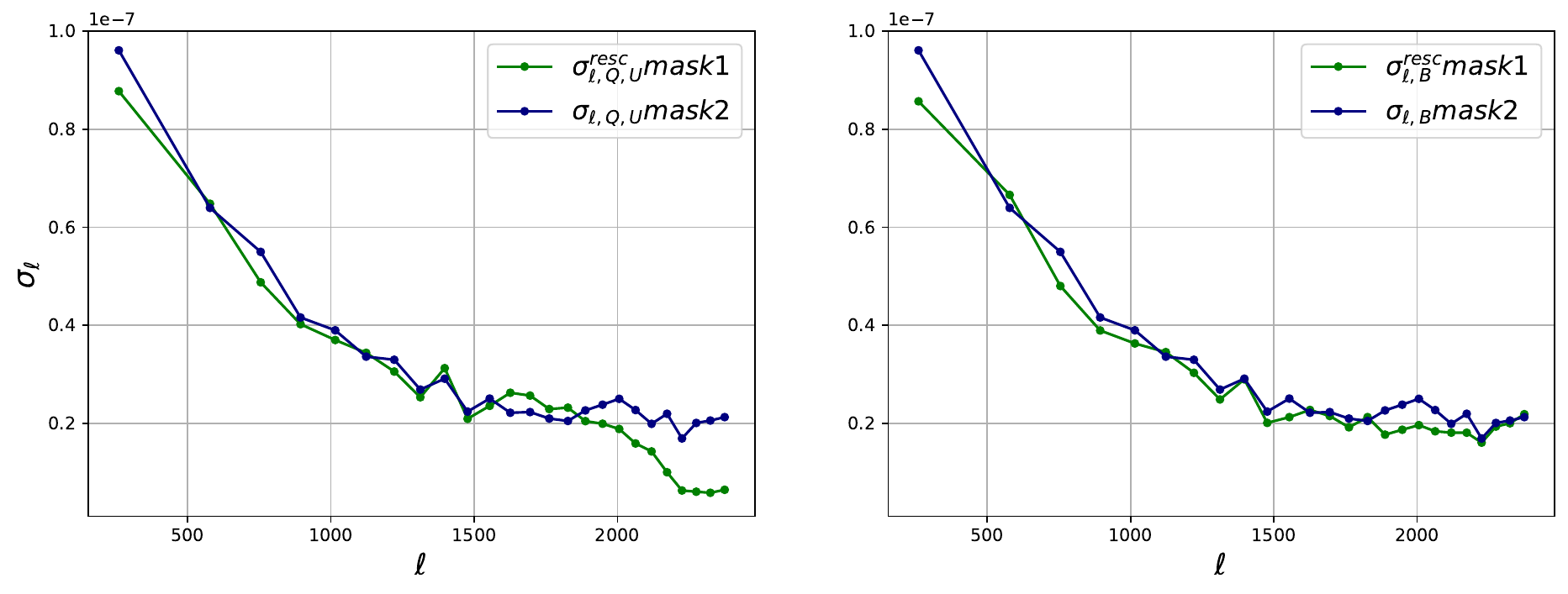}
\caption{Left panel: error in the power spectrum estimation for models trained with $J_{Q,U}$~\eqref{pol} and data with Mask1 (rescaled by a factor $\sqrt{f_{sky,mask1}/f_{sky,mask2}}$) and Mask2 applied. Right panel: error in the power spectrum estimation for models trained with $J_{B}$~\eqref{polB} and data with Mask1 (rescaled by a factor $\sqrt{f_{sky,mask1}/f_{sky,mask2}}$) and Mask2 applied.}
\label{desv_mask1_mask2}
\end{figure}

As it can be notice from Figure~\ref{desvcl_mask1_mask2}, the errors on the left panels (data with Mask1) are smaller than the errors on the right panels (data with Mask2). Therefore, Figure~\ref{desv_mask1_mask2} compares the power spectrum errors for data with Mask1 and Mask2 applied, where for the Mask1 case the uncertainties are rescaled by a factor equal to $\sqrt{f_{sky,mask1}/f_{sky,mask2}}$, $f_{sky}$ representing the fraction of the sky that remains unmasked. The left panel shows the results using models trained with $J_{Q,U}$, while the right panel presents the results for models trained with $J_{B}$. 

It is clear that the errors are larger when a more extensive mask is applied, as the fraction of the sky unmasked becomes smaller, making the power spectrum estimation more challenging. Then, the primary difference between the uncertainties, in each case, comes from the extension of the mask, as when one of them is rescaled by this effect, the curves closely matches. In the next section, we aim to compare these errors with those obtained using the pseudo-$C_{\ell}$ approach.

\subsection{Comparison with pseudo-$C_{\ell}$ estimator}
\label{sec:namaster}

The pseudo-$C_{\ell}$ algorithm estimates the power spectrum of spin-0 or spin-2 fields using the public library \textsc{NaMaster}, introduced in~\citep{2019MNRAS.484.4127A}. This framework provides essential tools for power spectrum estimation of CMB maps, both on the sphere and in the flat-sky approximation, including contamination deprojection and $E/B$ purification. The key difference compared to the optimal quadratic estimator, described in Section~\ref{sec:methodology}, lies in the substitution of the inverse covariance matrix (expressed as $C^{-1}$ in equation~\eqref{optimal}) with its diagonal, making it a more computationally efficient method. This approximation is optimal when the data have uncorrelated pixels, which is true if the noise is large and uncorrelated, or if the underlying power spectrum is close to white.

Given the widespread use of \textsc{NaMaster}, we aim to compare the uncertainties computed with this library to the errors presented in the previous section, which applies the Wiener Filter through neural networks and calculates the optimal quadratic estimator via simulations. For that purposes, we estimate the \textit{true} power spectrum given a dataset with inhomogeneous noise and the different masks applied, with the pseudo-$C_\ell$ method. 

We follow the steps for the flat-sky approximation specified in~\citep{2019MNRAS.484.4127A}. The process begins with a naive estimator by calculating the auto-spectra of the observed map, which couples different multipoles $\ell$ due to the incomplete sky coverage. Consequently, it calculates an analytical expression for the mode-coupling matrix $M_{\ell\ell'}$, in order to correct this bias, based on~\citep{Kogut_2003}. Then, the pseudo-$C_{\ell}$ expression for spin-2 fields will be: 
\begin{equation}
\left(
    \begin{matrix}
         \tilde{C}^{EE}_{\ell}\\
         \tilde{C}^{EB}_{\ell}\\
         \tilde{C}^{BE}_{\ell}\\
         \tilde{C}^{BB}_{\ell}
    \end{matrix}
\right)
=
\left(
\begin{matrix}
    M^{22}_{\ell\ell'}
\end{matrix}
\right)
\left(
    \begin{matrix}
         C^{EE}_{\ell'}\\
         C^{EB}_{\ell'}\\
         C^{BE}_{\ell'}\\
         C^{BB}_{\ell'}
    \end{matrix}
\right),
\end{equation}
where the elements of $M^{22}_{\ell\ell'}$ depend of the mask harmonic coefficients $w_{lm}$ by the auto-spectra of the mask (note that only for demonstration purposes we put the expression of the curved sky case): 
\begin{equation}
W_{\ell} = \sum_{m} w_{lm}w^{*}_{lm}.
\end{equation}

In general, is not possible to invert $M^{22}_{\ell\ell'}$ directly, therefore, the coupling matrix is binned into band-powers. The steps of the pseudo-$C_{\ell}$ calculation can be summarized in three stages: coupling the different multipoles by calculating the auto-spectra of the field, then binning into band-powers, and finally decoupling the band-powers with the inverted binned coupling matrix. 
%after binning into band-powers, and decoupling them. 

Since no Wiener Filter is applied in this method, we must estimate the noise bias by averaging the result of applying the pseudo-$C_{\ell}$ estimator to a large number of noise realizations (100 noise maps). %\textcolor{magenta}{aca puedo poner simplemente los resultados que daría aplicar estos pasos, sin purificación y compararlo con lo mio. Luego comparación con purificación.} 

To mitigate E-to-B leakage, a purification method is available, which offers an analytical approximation for isolating the pure B-mode component. In the derivation of this expression, a differential operator, denoted as $\bi{D}^{B}_{s}$, is defined. Applying $\bi{D}^{B}_{s}$ to a scalar field will give a pure B-mode. Then, the B-mode coefficients of a spin-2 field \textbf{P} can be expressed as: 
\begin{equation}
    \tilde{B}_{l} \equiv \int d\bi{\hat{n}} w(\bi{\hat{n}})(_{s}Y^{B}_{l}(\bi{\hat{n}}))^{\dag} \bi{P} = \int d\bi{\hat{n}} w(\bi{\hat{n}})(\bi{D}^{B}_{s}Y_{l})^{\dag}\bi{P}(\bi{\hat{n}}), 
\end{equation}
where $w$ is the mask vector, that should be on the right of $\bi{D}_{s}^{B}$ in order to consider $\bi{D}^{B}_{s}(wY_{l})$ as a B-mode. Therefore, the pure B-mode component is: 
\begin{equation}
B^{p}_{l} = \int d\bi{\hat{n}}(\bi{D}^{B}_{s}(wY_{l}))^{\dag}\bi{P}(\bi{\hat{n}}).
\end{equation}

Finally, expanding $\bi{D}^{B}_{s}(wY_{l})$ results in an expression of $B^{p}_{l}$ that depends of the first and second derivatives of the mask vector, requiring apodization with a window function for accurate application. We refer to~\citep{2019MNRAS.484.4127A} for more details in the calculations. 

For both masks, we applied a Gaussian window function to smooth the edges and enable the use of the purification method provided by the library. The estimator, using purification, will be unbiased if the mask is sufficiently smooth and differentiable. It is not always straightforward to do that, since a fraction of the signal is lost by the apodization of the mask. 
%Tiene sentido comparar con el caso sin purificacion?

\begin{figure}
    \centering
    \includegraphics[width=1.\columnwidth]{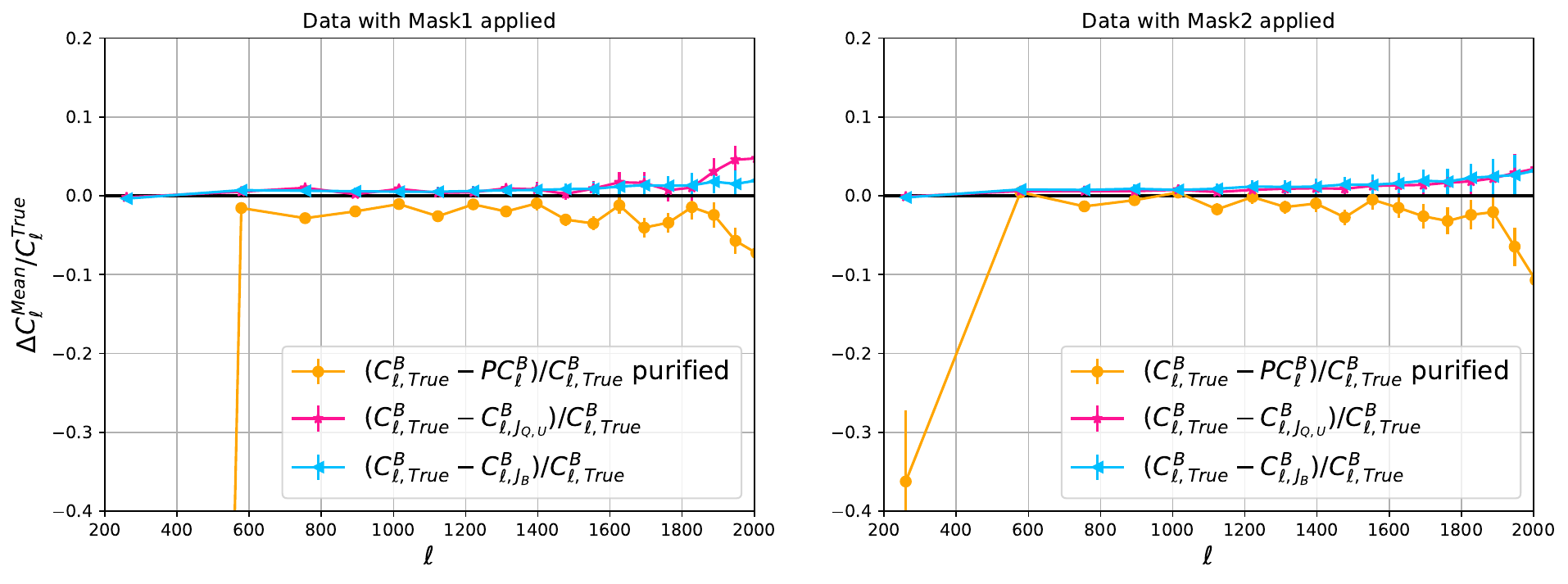}
    \caption{On the left panel it is presented the relative difference between the \textit{true} power spectrum and the average power spectrum estimation over 100 maps for data with Mask1 applied, while on the right panel it is presented the same comparison but for data with Mask2 applied.}
    \label{delta_namaster}
\end{figure}

We estimated the power spectrum for 100 maps and averaged the results. Figure~\ref{delta_namaster} presents the relative difference between the \textit{true} power spectrum and the average estimation obtained using the pseudo-$C_{\ell}$ calculation and our optimal quadratic estimator approach, along with their respective errors. 

It is evident that the average pseudo-$C_{\ell}$ estimation is slightly biased, especially in the first bin, and more pronounced for data with Mask1 applied (left panel), which does not occur with our estimation approach. Additionally, our optimal quadratic estimator method did not require the application of an apodization function.

\begin{figure}
    \centering
    \includegraphics[width=1.\columnwidth]{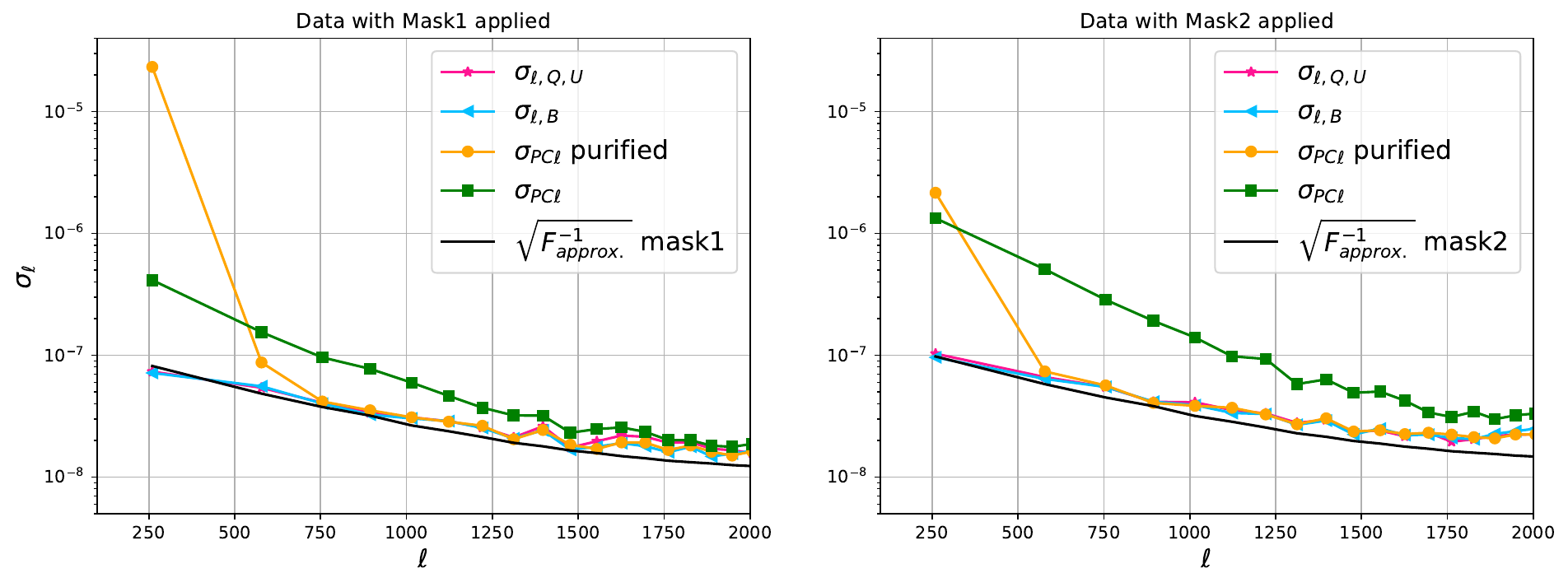}
    \caption{Comparison between the pseudo-$C_{\ell}$ estimation error (with and without purification) and the estimated errors following our approach. On the left panel for data with Mask1 applied, and on the right panel for data with Mask2 applied. Additionally, it is presented an approximated expression of the Fisher matrix (black curves).}
    \label{errores_namaster}
\end{figure}

In order to establish the precision using both methods, we compared the error of the power spectrum estimation for a single measurement. Figure~\ref{errores_namaster} presents the pseudo-$C_{\ell}$ error with $E/B$ purification in the orange curve, which is noticeably larger in the first bin (where the primordial B-modes appear) compared to the power spectrum error obtained with our approach. The second bin also exhibits a larger error than ours, but for scales beyond $\ell \approx 750$, the errors become quite similar. Therefore, the purification method results in smaller errors for scales beyond $\ell \approx 750$, but incurs inaccuracies in the largest scales. 

In addition, the green curves show the measurement errors without the purification method applied, which are larger than our uncertainties across all scales. It is clear that the $E/B$ purification reduces errors in the pseudo-$C_{\ell}$ estimation for scales beyond $\ell \approx 750$, but it performs worse at the largest scales, likely due to apodization.

At low $\ell$, in the first bin, our method significantly reduces the error bars compared to the pseudo-$C_{\ell}$, achieving a 99\% reduction for Mask1 and 95\% for Mask2 when considering the purified pseudo-$C_{\ell}$. Without the purification method, the error bars are tightened by 85\% for Mask1 and 91\% for Mask2 in the first bin.

We have used, as reference, the approximation of the Fisher matrix when homogeneous noise is applied (considering the average of the inhomogeneous noise), $F_{approx} = f_{sky}F_{\ell\ell'}$, where $f_{sky}$ is the fraction of the sky unmasked (different for both masks), and $F_{\ell\ell'}$ is the cosmic variance error for maps without mask applied (without mode coupling).

Finally, as expected, the errors obtained using a quadratic estimator approach (with an approximated WF through neural networks) are smaller than the pseudo-$C_{\ell}$ errors, and the average estimation is slightly more accurate than the pseudo-$C_{\ell}$ estimation. Using DeepWiener for the WF simulation allows the efficiently application of the optimal quadratic estimator for the power spectrum computation given a set of noisy maps.

\section{Conclusion}
\label{sec:conclusion}

% Acà hacer un resumen de lo que hiciste, indicando las dos maneras de entrenar las redes (las dos J's posibles). Puede ser en un solo parrafo que ayude a fijar la idea, preesentada de forma resumida.        
%
%despues continuar con lo que se hizo en el resto del paper, y la contribución principal de este trabajo. Mas las open questions para el futuro.

In this paper, we present a neural network model called DeepWiener, designed to simulate the Wiener Filter for application to CMB polarization maps with inhomogeneous noise applied. Using the trained models, we efficiently compute the E-mode and B-mode power spectra, implementing a simulation-based optimal quadratic estimator within a reasonable computation time.

The neural network’s performance is highly dependent on careful hyperparameter selection; however, achieving optimal reconstruction for both E-modes and B-modes is challenging due to E-to-B leakage. Usually, E-modes are well-reconstructed in the initial training, as they dominate the $Q$ and $U$ maps. To address this, we propose an iterative approach: in each iteration, a new dataset is generated with the E-mode reconstruction from the previous iteration removed, allowing the network to capture the B-modes in the signal.

We employed the loss function $J_{Q,U}$ in equation~\eqref{pol} when the network's inputs are the $Q$ and $U$ maps, which is the same expression as the $\chi^{2}$ function that must be minimized to obtain the WF solution. When the dataset predominantly contains B-modes, with most E-mode contribution removed, we also apply the proposed loss function $J_{B}$ in equation~\eqref{polB}, similar to the case of a scalar field like temperature. Here, the neural network receives the observed B-mode as input and directly outputs the WF of the B-mode map.

In Figure~\ref{rle}, we demonstrate that the reconstruction of the E-mode improves with each iteration using the loss function $J_{Q,U}$, as it also recovers residual E-mode contributions in the $Q$ and $U$ maps. Figure~\ref{rlb} shows the B-mode reconstruction in the final iteration, with a comparison of both loss functions $J_{Q,U}$ and $J_{B}$. It is evident that the B-mode WF obtained using the loss function $J_{B}$ aligns more closely with the WF produced by the PCG method, than the reconstruction with the loss function $J_{Q,U}$. This likely occurs because the neural network trained with $Q$ and $U$ maps attempts to recover some residual E-modes, while the network receiving the observed B-mode map focuses exclusively on that component. The use of the loss function $J_{B}$ is appropriate once the E-to-B leakage is significantly reduced; otherwise, the observed B-map would be heavily contaminated by E-mode contributions.

The cases considered include CMB polarization maps with inhomogeneous noise applied, and two masks, covering different fraction of the sky. Given that the primary challenge in this work is the E-to-B leakage, we demonstrate the neural network's capacity to simulate the WF across datasets with different mask complexities. For a dataset with Mask1 applied (see Figure~\ref{mascaras}), 4 iterations were necessary, whereas 5 iterations were required for Mask2, using loss function $J_{Q,U}$. When using loss function $J_{B}$, the number of required iterations decreased to 3 for Mask1 and 4 for Mask2.

Additionally, we present the power spectrum estimation of an unknown signal for the E-modes and B-modes using a simulation-based optimal quadratic estimator. This approach was feasible because predictions made with the trained models are significantly faster than computing the WF via the PCG method, even though each map requires 4 or 5 model applications. In Appendix~\ref{apx4} we calculate the computation time for DeepWiener models compared with PCG. Calculating the noise bias required applying the WF to 100 maps, and the Fisher matrix calculation to 2000 maps, a scale that would be impractical to achieve with the traditional Conjugate Gradient method within a reasonable time. 

To verify that the procedure was performed correctly, Figure~\ref{desvcl_mask1_mask2} shows that the errors in the power spectrum estimation closely match the square root of the inverse of the Fisher matrix, which, by definition, represent the covariance matrix of the power spectrum. 

We remark that achieving satisfactory convergence for the B-mode reconstruction is more challenging for the dataset with Mask2 applied compared to Mask1. Furthermore, Figure~\ref{desv_mask1_mask2} clearly shows that the power spectrum errors are higher for the dataset with Mask2 due to the smaller unmasked sky fraction.

Finally, we show on Figures~\ref{delta_namaster} and~\ref{errores_namaster} that the estimation with our procedure is more accurate than the pseudo-$C_{\ell}$ method, as our errors are smaller in the first bins and the average estimation is unbiased. It is expected since the pseudo-$C_{\ell}$ method does not solve the optimal quadratic estimator.

In future work, we aim to apply these methods on a real-world CMB experiments, such as the Q $\&$ U Bolometric Interferometer for Cosmology, QUBIC~\citep{Hamilton2022}, with more complex noise properties. Besides, we plan to explore the improvements in the neural network architecture by incorporating more complex Deep Learning models currently used in Computer Vision~\citep{2020arXiv201011929D}. Another possible approach would be to implement a new neural network architecture that directly estimates the power spectrum from a set of observed maps, using a quadratic form in the data, similar to the optimal quadratic estimator. 

%%%%%%%%%%%%%%%%%%%%%%%%%%%%%%%%%%%%%%%%%%%%%%%%%%%%%%%%%%%%%%%
\appendix

\section{Conjugate gradient algorithm}
\label{apx1}

The conjugate gradient (CG) is an iterative numerical method that solve linear system equations of the form: 
\begin{equation}
    \textbf{A}\textbf{x} = \textbf{b},
\label{CG}
\end{equation}
where the known $n \times n$ matrix $\textbf{A}$ is real, symmetric ($\textbf{A}^{T}=\textbf{A}$) and positive definite ($\textbf{x}^{T}\textbf{Ax} > 0$, for all non-zero $\textbf{x}$).

In order to find the WF estimator, the system that needs to be solved is: 
\begin{equation}
   (\textbf{S}^{-1}+\textbf{R}^{\dag}\textbf{N}^{-1}\textbf{R})\textbf{x} = \textbf{R}^{\dag}\textbf{N}^{-1}\textbf{d}, 
\end{equation}
where $\textbf{d}$ represents the $Q$ and $U$ components in configuration space,
$\textbf{x}$ corresponds to the $E$ and $B$ components in Fourier space,
and the response matrix $\textbf{R}$ acts as an operator that transforms from Fourier space to configuration space and applies the rotation from $Q$ and $U$ to $E$ and $B$, as detailed in Appendix~\ref{apx2}.
%
%The goal is to obtain the inverse of the matrix $(\textbf{S}^{-1}+\textbf{R}^{\dag}\textbf{N}^{-1}\textbf{R})$,
The goal is to invert the problem and find a solution of the unknown vector $\textbf{x}$. 
%iteratively, starting from an initial guess $\textbf{x$_0$}$.
%of the unknown vector $\textbf{x}$. 
Compared to equation~\eqref{CG}, in this case, $\textbf{b} = \textbf{R}^{\dag}\textbf{N}^{-1}\textbf{d}$ and $\textbf{A} = (\textbf{S}^{-1}+\textbf{R}^{\dag}\textbf{N}^{-1}\textbf{R})$. 

The general idea of the algorithm is to start from an initial guess for $\textbf{x}$, $\textbf{x}_{0}=0$, which is iteratively updated based on a metric indicating its proximity to the solution $\textbf{x}_{*}$. The solution has to be the unique minimizer of the quadratic function: 
\begin{equation}
    f(\textbf{x}) = \frac{1}{2} \textbf{x}^{T} \textbf{A x} - \textbf{x}^{T} \textbf{b}, \hspace{1cm} \textbf{x} \in R^{n}.
\end{equation}

This function is minimized when its gradient is equal to zero: 
\begin{equation}
    \triangledown f = \textbf{Ax} - \textbf{b}, 
\end{equation}
which is equal to equation~\eqref{CG}. The minimization proceeds by generating a sequence of search directions $\textbf{p}_{k}$, and updated solutions, $\textbf{x}_{k}$. In each iteration, a quantity $\alpha_{k}$ is determined that minimizes $f(\textbf{x}_{k}+\alpha_{k}\textbf{p}_{k})$, and $\textbf{x}_{k+1}$ is then set to $\textbf{x}_{k}+\alpha_{k}\textbf{p}_{k}$.

The algorithm begins by defining the residual  $\textbf{r} = \textbf{b}-\textbf{Ax}$. If the residual is sufficiently small for the initial guess $\textbf{x}_{0}$, then $\textbf{x}_{0}$ is returned as the result. Otherwise, set $\textbf{p}_{0}=\textbf{r}_{0}$ and enter the iteration loop, which proceeds as follows:
\begin{align}
    \alpha_{k} = & \frac{\textbf{r}^{T}_{k}\textbf{r}_{k}}{\textbf{p}^{T}_{k}\textbf{A}\textbf{p}_{k}} \\
    \textbf{x}_{k+1} = & \textbf{x}_{k} + \alpha_{k}\textbf{p}_{k} \\
    \textbf{r}_{k+1} = & \textbf{r}_{k} - \alpha_{k}\textbf{Ap}_{k} \\
    \beta_{k} = & \frac{\textbf{r}^{T}_{k+1}\textbf{r}_{k+1}}{\textbf{r}^{T}_{k}\textbf{r}_{k}} \\
    \bi{p}_{k+1} =  & \textbf{r}_{k+1} + \beta_{k}\textbf{p}_{k},
\end{align}
The loop continues until the residual $\textbf{r}_{k+1}$  is sufficiently small, at which point  $\textbf{x}_{k+1}$ is returned as the result. For a more detailed explanation of the procedure, refer to~\citep{10.5555/1403886}.

The matrix $(\textbf{S}^{-1}+\textbf{R}^{\dag}\textbf{N}^{-1}\textbf{R})$ can become ill-conditioned when its condition number is large, which can significantly slow down the method.  This poor conditioning is typically caused by $\textbf{S}^{-1}$.  To improve the conditioning, it is necessary to define a preconditioner, which is a non-singular matrix that transforms the problem into an equivalent one with better conditioning. A simple choice for the preconditioner is $\textbf{S}^{-1}$~\citep{Chen_2005}. 

Following the described method, we developed our own code to solve the linear system, based on the \textsc{quicklens} package\footnote{\url{https://github.com/dhanson/quicklens}}, which is included in the DeepWiener repository.

%, \textcolor{magenta}{that can be seen on the script called "CG_inho.py".}

\section{Partial sky polarization}
\label{apx2}

The CMB radiation is characterized by the Stokes parameters $I$, $Q$, and $U$. The parameter $I$ denotes the intensity of the radiation at an observed position $\hat{n}$, while $Q$ and $U$ describe the linear polarization in Cartesian coordinates, rotated by $45^\circ$, respectively. The CMB temperature anisotropies are measured as fluctuations in intensity, which correspond to a spin-0 (scalar) field. In contrast, the linearly polarized components $Q$ and $U$ are coordinate-dependent quantities that transform under coordinate rotations in the plane perpendicular to direction $\hat{n}$. Then, the polarization can be expressed using two complex spin-2 fields: 
\begin{equation}
    P_{\pm} = Q \pm i U, 
\end{equation}
which transform as: 
\begin{equation}
    P_{\pm} \mapsto   P_{\pm} e^{\mp2i\phi},
\end{equation}
under a right-handed rotation by an angle $\phi$.

These fields can be decomposed using the spin-weighted spherical harmonics $_{s}Y_{\ell m}$ with spin weights $s = \pm 2$: 
\begin{equation}
  P_{\pm}(\hat{n}) = \sum_{\ell m}a_{\pm 2,\ell m} Y_{\pm 2,\ell m}(\hat{n}),  
\end{equation}
where $\hat{n}$ is the observational position vector and $a_{\pm 2,\ell m}$ are the expansion coefficients. 

It is convenient to express the polarization in terms of the scalar E-mode and pseudo-scalar B-mode, to define the power spectrum using rotationally invariant quantities~\citep{1997PhRvD..55.1830Z,1997PhRvD..55.7368K}, where the harmonic coefficients are: 
\begin{align}
    a_{E, \ell m} & = -\frac{1}{2}(a_{2,\ell m} + a_{-2,\ell m}), \\
    a_{B, \ell m} & = \frac{i}{2}(a_{2,\ell m} - a_{-2,\ell m}).
\end{align}

These combinations behave differently under parity transformation: 
\begin{align}
    a_{E, \ell m} & \mapsto (-1)^{\ell} a_{E, \ell m}, \\
    a_{B, \ell m} & \mapsto (-1)^{\ell + 1} a_{B, \ell m}
\end{align}

In the flat-sky approximation, instead of spherical decomposition, a plane wave expansion is used. For example, for temperature anisotropies: 
\begin{equation}
    \sum_{\ell m}a_{T,\ell m} Y_{\ell m} (\bi{\theta}) \rightarrow  \int d^{2}\bi{l}T(\bi{l})e^{i\bi{l}\cdot \bi{\theta}},
\end{equation}
where, instead of the multipoles $\ell, m$, the analysis is performed in terms of $\bi{l}$, a vector in the 2D Fourier plane. Then, the definitions of the E and B-modes are modified accordingly~\citep{2000PhRvD..62d3007H}: 
\begin{align}
    a_{E, \ell m} & \rightarrow E(\bi{l}) \equiv Q(\bi{l})cos(2\phi_{l})+U(\bi{l})sen(2\phi_{l}), \\
    a_{B, \ell m} & \rightarrow B(\bi{l}) \equiv -Q(\bi{l})sen(2\phi_{l})+U(\bi{l})cos(2\phi_{l}), 
\end{align}
where $\phi_{l}$ is the angle between $\bi{l}$ and the positive $l_{x}$ axis. %\textcolor{magenta}{pongo más cuentas intermedias?}

\section{Neural network iterations}
\label{apx3}

As referenced throughout the paper, especially in the results section~\ref{sec:results}, we performed several training stages, specified as iterations, where the E-mode contribution from the previous iteration is progressively removed. This iterative approach is needed because the CMB polarization is predominantly composed of E-modes, with amplitudes at least two orders of magnitude larger than B-modes. Nevertheless, although the primordial B-modes have a very faint signal, developing new techniques to detect them is crucial, as they serve as evidence of the inflationary epoch.  

The observed $Q_{obs}$ and $U_{obs}$ polarization maps, that an instrument could measure, contain contributions from both the E-modes and B-modes. Since the E-modes represent the dominant signal in these maps, the first training of DeepWiener extract the polarization information present on these modes. While the reconstructed E-mode signal, $E_{NN}$, obtained by the neural network is generally accurate, the main objective is to recover the B-mode signal, $B_{NN}$, which is largely obscured by the E-mode contribution.

In each iteration, a new dataset is generated by removing the $E_{NN}$ contribution obtained from the previous iteration, enabling the network to focus on B-modes, which are the remaining contribution in the polarization maps. The outputs of DeepWiener, depending of the inputs provided, yield both a correction to the E-mode reconstruction from the previous iteration and the extracted B-mode contribution for the current iteration. 

\begin{figure}
    \centering
    \includegraphics[width=1.\columnwidth]{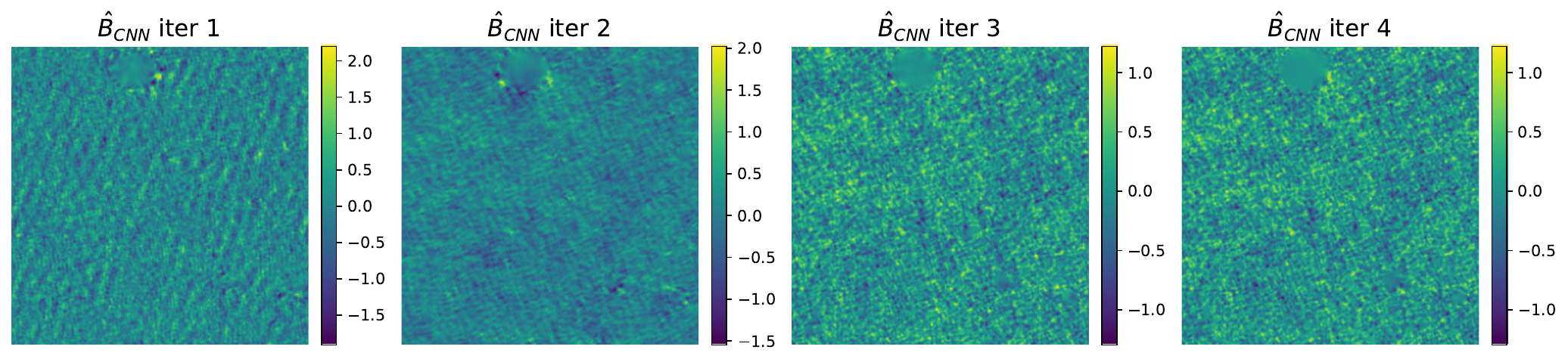}
    \caption{B-mode reconstruction in each DeepWiener iteration, for a dataset with Mask1 applied and using the loss function $J_{Q,U}$, equation~\eqref{pol}.}
    \label{iteraciones_bwf}
\end{figure}

Figure~\ref{iteraciones_bwf} presents the WF of the B-modes for each iteration, obtained with DeepWiener, for the dataset with Mask1 applied. Each iteration shows an improvement in the B-mode reconstruction. The first iteration fails to recover almost any B-mode signal, while the final iteration closely resembles the expected result from the PCG algorithm.

\begin{figure}
    \centering
    \includegraphics[width=1.\columnwidth]{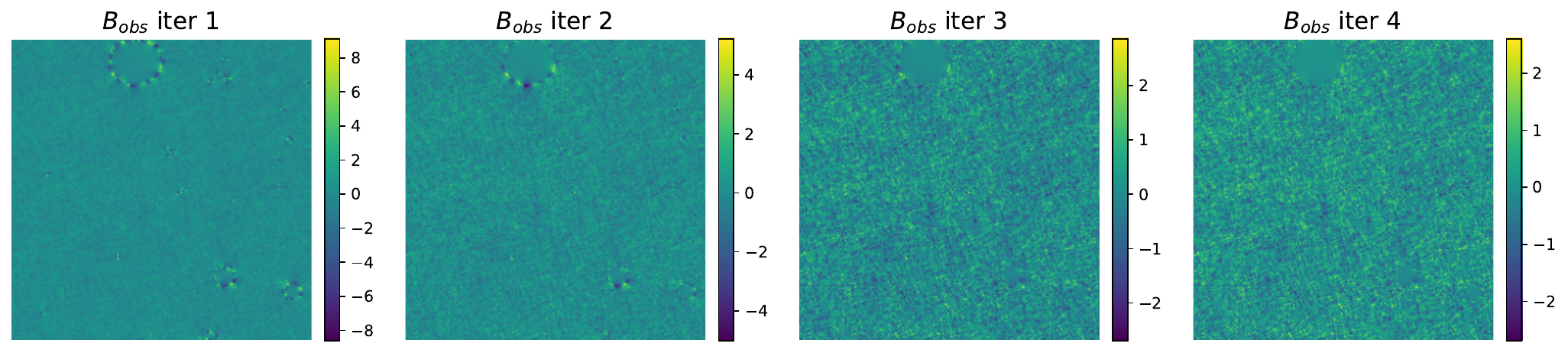}
    \caption{Observed B-mode transformed from $Q^{i}_{obs}$ and $U^{i}_{obs}$ at each iteration, for a dataset with Mask1 applied.}
    \label{iteraciones_bobs}
\end{figure}

From Figure~\ref{iteraciones_bobs} it is evident that the observed B-mode map, transformed from $Q^{i}_{obs}$ and $U^{i}_{obs}$, progressively reduces the E-to-B leakage at the edges of Mask1. This improvement occurs as the E-mode contribution is incrementally removed with each iteration. The observed B-mode map of the third iteration is used as input of DeepWiener when it is trained with the loss function $J_B$, equation~\eqref{polB}.

\section{Computation time}
\label{apx4}

In previous work~\citep{2023BAAA...64..193C, 2024JCAP...04..041C}, we investigated the computational efficiency of applying the WF using neural network models compared to the PCG method. We tested various cases involving temperature maps with homogeneous noise, considering different noise levels and number of pixels. The results demonstrated that once the neural network is trained, performing predictions on a test set is significantly faster and more computationally efficient than calculating the WF through the PCG algorithm.

%A neural network is trained only once, and the resulting trained model can then be applied to any map with the same characteristics as those used during training.
The proposed neural network architecture is trained with two GPU A100, with a training time of $\thicksim$ 18 h over 200 epochs and using a training dataset of 5000 maps. As it is necessary to train several models to reach the expected accuracy in the B-mode map, the training time could increase to 3 days in the most demanding scenario (with 5 models for the Mask2). It is important to highlight that the training of the neural network is conducted only once, then the resulting trained model can then be applied to a dataset of any size, containing maps with the same characteristics as those during training.

Because the predictions with the trained models are extremely fast, this approach allows the WF to be applied to a large number of maps efficiently. Conversely, implementing the WF with the PCG method requires performing iterative inversion for each map. This makes the PCG approach computationally inefficient and even prohibitive when working with large maps.

Having a faster method to approximately perform the WF, enable to calculate the noise bias term and fisher matrix through simulations (100 maps for the noise bias and 2000 maps for the fisher matrix). 

In this work, when new $Q_{obs}$ and $U_{obs}$ maps are received, it is necessary to apply a series of trained models depending on the mask used and the loss function implemented. For instance, using the loss function $J_{Q,U}$, 5 models were used for Mask2 and 4 models for Mask1. When the loss function $J_{B}$ is used, 4 models were necessary to achieve a good performance considering the Mask2, and 3 models for the Mask1. It is important to note that the prediction time will depend on the number of weights of the model, which is the same for all of them in any Mask case. Then the computation time will only increase if more models are implemented, independently of the Mask and loss function used.  

Then, we calculate the computational time to apply those models on a test set of 100 maps and compare it with the computational time with the PCG algorithm. The last method will depend on the complexity of the Mask used and, therefore will be different for Mask1 and Mask2.

These results are presented in Table~\ref{tabla_time_CNN_CG}, in seconds and minutes, where it can be noticed that the prediction time using several models is faster (a factor order of 10) than the PCG method. Note that these calculations were performed on a CPU.

Although this comparison in computational time is based on 100 maps, the Fisher matrix estimation requires applying the WF to 2000 maps or more, where the PCG method could take several days to complete.

\begin{table}[ht!]
\centering
\begin{tabular}{|c| ccc| cc | }
%\hline\hline\noalign{\smallskip}
\hline
{} & {} &  CNN & {}  &  PCG  & {}  \\
\hline
Time & 3 models & 4 models & 5 models & Mask1 & Mask2  \\
%\hline\noalign{\smallskip}
\hline
[sec] & 462.85  & 628.55 & 792.19 & 6776.4 & 9756.2 \\
\hline
[min] & 7.71  &  10.4 & 13.20 &  112.94 & 162.60 \\
\hline
\end{tabular}
\caption{Computing time, in seconds and minutes, required to estimate the WF on 100 maps. Left table: using CNN and different number of models depending of the case.  Right table: using PCG and different Mask applied to the data. }
\label{tabla_time_CNN_CG}
\end{table}

\section*{Acknowledgements}

M.B.C. acknowledges a doctoral fellowship by CONICET.
C.G.S. and M.B.C. acknowledge funding from CONICET (PIP-2876), and Universidad Nacional de La Plata (G11-175), Argentina. 
M.Z. is supported by NSF 2209991 and NSF-BSF 2207583.
The training of the neural networks and the simulations used here were performed at the SNS Supercomputing Center, at the Institute for Advanced Studies, in Princeton.

%%%%%%%%%%%%%%%%%%%% REFERENCES %%%%%%%%%%%%%%%%%%

%% [A] Recommended: using JHEP.bst file
 \bibliographystyle{JHEP}
 \bibliography{bibliografia_DeepWiener.bib}

\end{document}